\documentclass[twocolumn,trackchanges]{aastex701}
\usepackage[utf8]{inputenc}
\usepackage{amsmath}
\usepackage{multirow}
\usepackage{hyperref}
\usepackage{graphicx} 
\usepackage{booktabs}
\usepackage{subcaption} 
\usepackage{color}
\usepackage{algorithm}
\usepackage{siunitx}
\usepackage{algorithmicx}
\usepackage{algpseudocode}

\newcommand{\halpha}{H$\alpha$}

\newcommand{\curf}{\mathcal{F}}
\usepackage[T1]{fontenc}

\usepackage{scalerel}
\newcommand{\CaII}{{\ion{Ca}{2}}}

\newcommand{\CaIIk}{{\ion{Ca}{2}} K}

\newcommand{\av}{A$_{\text{V}}$}
\newcommand{\mdot}{$\dot{\text{M}}$}
\newcommand{\Mdot}{{\dot{{M}}}}

\newcommand{\msun}{ M_{\sun}}
\newcommand{\rsun}{ R_{\sun}}

\newcommand{\msunyr}{M_{\sun} \, \rm{ yr^{-1}}}

\newcommand{\teff}{T$_{\rm eff}$}
\newcommand{\logg}{$\log g$}
\newcommand{\feh}{[Fe/H]}
\newcommand{\vsini}{$v\sin i$}
\newcommand{\rotfit}{{\tt ROTFIT}}

\newcommand{\tmax}{T$_{\rm max}$}

\definecolor{dgreen}{rgb}{0.1, 0.53, 0.22}
\definecolor{bluee}{rgb}{0.03, 0.42, 0.84}

\begin{document}

\title{Connecting Calcium Depletion and Dust Evolution in the Rapid Accretor T28}
\correspondingauthor{Eylul Galimidi}

\author[0009-0000-0630-5735]{Eylul Galimidi}
\affiliation{Department of Astronomy, University of Michigan, 1085 South University Avenue, Ann Arbor, MI 48109, USA.}
\email[show]{eylul@umich.edu}

\author[0000-0002-3950-5386]{Nuria Calvet}
\affiliation{Department of Astronomy, University of Michigan, 1085 South University Avenue, Ann Arbor, MI 48109, USA.}
\email{ncalvet@umich.edu}

\author[0000-0001-8022-4378]{Marbely Micolta}
\affiliation{Department of Astronomy, University of Michigan, 1085 South University Avenue, Ann Arbor, MI 48109, USA.}
\email{micoltam@umich.edu}

\author[0000-0002-0474-0896]{Antonio Frasca}
\affiliation{INAF – Osservatorio Astrofisico di Catania, via S. Sofia 78, 95123 Catania, Italy}
\email{antonio.frasca@inaf.it}

\author[0000-0001-6647-862X]{Ezequiel Manzo-Martínez}
\affiliation{Department of Astronomy, University of Michigan, 1085 South University Avenue, Ann Arbor, MI 48109, USA.}
\affiliation{Instituto de Astronomía, Universidad Nacional Autónoma de México, Km. 107 Carretera Tijuana-Ensenada, Ensenada Baja California, México, CP 22860, Mexico}
\email{emanzo@umich.edu}

\author[0000-0001-7124-4094]{César Briceño}
\affiliation{SOAR Telescope/NSF NOIRLab, Casilla 603, La Serena, Chile}
\email{cesar.briceno@noirlab.edu}


\begin{abstract}

\noindent
We present a multi-wavelength case study of the star T28 in the Chamaeleon I star-forming region and investigate
the dust evolution of its highly Ca-depleted protoplanetary disk. We combine new observations from 
the MIKE spectrograph on the
Magellan Clay Telescope  with archival IRS spectrum and photometry.
We constrain the optical extinction using the hydrogen column density of the region, and apply \rotfit{} to obtain stellar parameters. 
We estimate the mass accretion rate by fitting the excess flux above the photosphere with an accretion shock model. We model the line profiles of \CaIIk{} and H$\alpha$ with magnetospheric accretion models to infer Ca/H abundance, while the infrared excess is modeled with
irradiated accretion disk models to get disk properties. We find that T28 is a high accretor with significant Ca depletion in its accretion flows.
We find that the disk is best fit with a \textit{pre-transitional} model with a disk mass of 0.12 $\msun{}$, where two optically thick disks are separated by a gap filled with optically thin dust extending to 16 au. This gap would not be resolved in the existing observations. T28 follows the trend of increasing depletion with increasing mass accretion rate. This trend can potentially be explained by preferential and early giant planet formation in massive disks, as suggested by recent planet formation simulations, leading to enhanced capture of refractory material in the massive disks of high accretors and thus depletion in the material reaching the star.

\end{abstract}

\section{Introduction} \label{intro}
Young stars are born surrounded by gas-rich disks, which evolve into planetary systems. Although much progress in understanding the processes involved in this evolution has been made in recent years, many details of how these processes take place are still to be understood.


In disks no longer embedded in the cloud from which they formed, the inherited interstellar medium (ISM) dust, which constitutes about 1\% of the primordial disk mass, has already grown and settled towards the midplane to varying degrees \citep{2004A&A...421.1075D, dalessio_effects_2006, 2006ApJS..165..568F, 2020ApJ...893...56M}. Most of the mass in solids is 
bound in particles of millimeter/centimeter size, called pebbles. The refractory material in these pebbles
remains bound until pebbles reach the dust sublimation front, at which the temperature is high enough to sublimate. Thus, the distribution of refractories throughout the disk traces the dust evolution and the effects of radial transport and trapping in the disk, which set the conditions for planet formation.

One such mechanism of dust evolution is radial drift, by which pebble-size particles move inwards in the disk faster than the gas \citep{weidenschilling_origin_1997,birnstiel_simple_2012}. This process leads to a brief period of refractory enrichment in the inner regions of the disk,
followed by depletion as matter is accreted onto the star \citep{huhn_how_2023, 2021A&A...654A..71S, 2021A&A...654A..72S}. During this process, planets in early stages of formation capture pebbles to build their cores \citep[e.g.][]{lambrechts_rapid_2012}. 
In addition, pressure bumps at the edges of
gaps in the disk induced by planets trap pebbles and
block the inward drift of material \citep[e.g.][]{Pinilla_2012, Zhu_2012, van_der_Marel_2021}, further reducing the abundance of rocky material that reaches the inner regions of the disk \citep{huhn_how_2023, 2021A&A...654A..71S, 2021A&A...654A..72S}.

In support of these theoretical predictions, \citet{micolta_ca_2023, Micolta_2024} from analyses of the Ca II K line of low mass stars (T Tauri stars, TTS), determined underabundances of calcium relative to solar, an element found in species with the highest condensation temperatures \citep{lodders_solar_2003}. Ca depletion is found in disks in the Chamaeleon and Lupus star-forming regions, with ages between 1 and 3 Myr, as well as in the 5 Myr Orion OB1b subassociation \citep{briceno2019}. Moreover, Ca depletion is found in all disks with reported substructures, including disks with large inner regions mostly void of small grains \citep[transitional,][]{2005ApJ...630L.185C} and disks with gaps \citep[pre-transitional,][]{2007ApJ...670L.135E}, as expected if refractory elements are captured in pressure traps or planets. Similar results have been found for other TTS \citep{McClure_2019,McClure_2020} and young intermediate mass stars \citep[Herbig Ae,][]{Kama_2015,2023A&A...671A.140G}.

\citet{micolta_vanishing_2025} analyzed refractory abundance in the TTS BP Tau, surrounded by a disk which originally had been classified as compact and lacking structure \citep{2019ApJ...882...49L}. \citet{micolta_vanishing_2025} determined abundances relative to H of calcium and of magnesium, another refractory element, finding that
both elements were depleted relative to solar. Notably, Mg was less depleted than Ca, as expected from the lower condensation temperature of Mg-bearing species \citep{lodders_solar_2003}. Concurrent with this work, reanalysis of previous ALMA data and new observations found a relatively small gap in the disk of BP Tau \citep[]{Zhang_2023, Yamaguchi_2024, 2025A&A...694A.147G}. The presence of a gap was required to explain the spectral energy distribution (SED) and was consistent with the refractory depletion found in BP Tau.
 
In this work, we study T28\footnote{Full SIMBAD resolvable name is Ass Cha T 2-28} (2MASS J11074366-7739411), a classical (accreting) T Tauri star (CTTS) in the Chamaeleon I star forming region. This star showed a strong depletion of Ca \citep{micolta_ca_2023}, surrounded by a compact disk \citep{luhman_disk_2008, pascucci_steeper_2016}.   
Previous attempts at characterizing this disk system have been largely limited due to uncertainties in the optical extinction $\text{A}_{\text{V}}$, which have resulted in values of A$_V$ differing by a factor of $\approx $ 2 \citep[]{luhman_stellar_2007, manara2017}. In this paper, we revisit the \av{} measurement and redetermine stellar, accretion, and disk properties. Our determination of the mass accretion rate for T28 is higher than previous determinations \citep{manara2017}. The mass accretion rate value is also about a factor of 20 higher than the accretion rate of BP Tau \citep{2013ApJ...767..112I}. As such, the analysis of T28 allows us to probe a different part of the parameter space than that covered in \citet{micolta_vanishing_2025}.

We perform a detailed analysis of the SED following \citet{mcclure_curved_2013} and \citet{mauco_herschel_2018}, and search for indications of substructure as expected from the high level of depletion. We investigate potential radial gradients in the relative abundances of refractories (e.g., Mg/Fe ratio), and check whether the chemical depletion trends found in BP Tau hold \citep{micolta_vanishing_2025}.

This paper is organized as follows: In Section \ref{data}, we outline the observations taken and the data used. In Section \ref{models}, we discuss the models used. In Section \ref{analysis}, we present our results and analysis. In Section \ref{discussion}, we discuss the implications and possible interpretations of our results. Finally, in Section \ref{summary}, we give our conclusions.

\section{Observations} \label{data}
\subsection{New Observations}

\subsubsection{Spectra}
We obtained an optical spectrum of T28 on 2026 January 28 with the Magellan Inamori Kyocera Echelle \citep[MIKE,][]{Bernstein_2003} spectrograph on the Magellan Clay telescope at the Las Campanas Observatory in Chile. We used the 0.7" slit and the blue and red cameras
covering a wavelength range of 3200 \AA\ to 10000 \AA,
with a spectral resolution of R $\approx$ 32,500 in the red and $\approx$ 40,000 in the blue. We used an exposure time of 4620 s for each camera. We reduced the data using the CarPy package \citep{2000ApJ...531..159K, kelson_optimal_2003}. 

We note that since MIKE is an echelle spectrograph, the observed spectrum is dominated by the blaze function, wavelength-dependent throughput, and slit losses. Although the reduction process removes the instrumental response, it does not preserve the absolute continuum shape. Consequently, MIKE does not recover broad continuum features. Instead, the overall continuum shape constrained through photometric calibration. As such, we do not recover a Balmer jump and interpreting the continuum shape below 3900 \AA\ requires caution.

\subsubsection{Concurrent Photometry}
We obtained $griz$ photometry within an hour of MIKE observations using the Goodman High Throughput Spectrograph red camera \citep[][]{clemens_goodman_2004} at the Southern Astrophysical Research (SOAR) telescope in the Cerro Tololo Inter-American Observatory (CTIO). We performed aperture photometry using The Goodman Spectroscopic Pipeline v2.0.0rc1 release\footnote{https://soardocs.readthedocs.io/projects/goodman-pipeline/en/latest/}. 
Table~\ref{tab:newphot} shows the magnitudes of T28 from new photometric observations with propagated errors.

\begin{deluxetable}{lcc}[t!]
\tablecaption{SOAR/Goodman photometry \label{tab:newphot}}
\tablehead{\colhead{Filter} & \colhead{$\lambda$ ($\mu$m)} & \colhead{$m \pm \sigma(m)$}}
\startdata
CTIO g& 0.468& 16.58 $\pm$ 0.56\\
CTIO r& 0.621& 15.31 $\pm$ 0.98 \\
CTIO i& 0.765 & 13.28 $\pm$ 0.26\\
CTIO z& 0.969& 12.85 $\pm$ 0.21\\
\hline
\enddata
\end{deluxetable}


\subsection{Ancillary Observations}\label{sec:ancillary} 
\subsubsection{Mid-Infrared Spectrum}
T28 was observed with the Spitzer InfraRed Spectrograph \citep[IRS,][]{houck_infrared_2004, werner_spitzer_2004} in 2005 as part of Program 2 (PI: Houck). We downloaded the IRS spectrum from the NASA/IPAC Infrared Science Archive (AORkey 12696832). 

\subsubsection{Photometry}
We obtained the photometry using the VizieR catalog access tool \citep{Ochsenbein_2000}. We include the Two Micron All Sky Survey (2MASS), Spitzer/IRAC, Herschel/PACS, as well as millimeter photometry from the Large APEX BOlometer CAmera \citep[LABOCA,][]{siringo_large_2009} on the APEX 12 m submillimeter telescope \citep{gusten_atacama_2006} and the Atacama Large Millimeter/Submillimeter Array \citep[ALMA,][]{brogan_2014_2015}. 
Table~\ref{tab:photometry} shows archival photometry. 

\subsubsection{WTTS}
The non-accreting weak-line T Tauri star (WTTS) TWA25 was observed with the X-shooter instrument on the Very Large Telescope (VLT) at the European Southern Observatory (ESO) \citep{2011A&A...536A.105V}. We used its spectrum and parameters from \citet{2013A&A...551A.107M} and take it to be the photospheric template for T28.

\begin{deluxetable*}{lcccccc}[t!]
\tablecaption{Photometry \label{tab:photometry}}
\tablehead{\colhead{Filter} & \colhead{$\lambda$ ($\mu$m)} & \colhead{$\lambda F_{\lambda}$} & \colhead{$\lambda F_{\rm err,\lambda}$} & \colhead{$\lambda F_{\lambda,0}$} & \colhead{$\lambda F_{\rm err,\lambda, 0}$}  & \colhead{Reference}}
\startdata
2MASS:J            & 1.24     & 3.3e-10&7.3e-12& 1.2e-9& 2.6e-11& 1\\
2MASS:H            & 1.65     & 4.9e-10&1.1e-11& 1.1e-9& 2.4e-11& 1\\
2MASS:Ks           & 2.16     & 4.7e-10&1.1e-11& 7.8e-10& 1.9e-11& 1\\
Spitzer:IRAC1& 3.55 & 2.7e-10&5.1e-12& 3.5e-10& 6.4e-12& 2 \\
Spitzer:IRAC2& 4.49     & 1.9e-10&3.3e-12& 2.2e-10& 3.9e-12& 2\\
Spitzer:IRAC3*& 5.73 & 1.1e-10&3.1e-12& 1.4e-10& 5.4e-12& 2\\
Spitzer:IRAC4*& 7.87 & 8.1e-11&3.0e-12& 1.0e-10& 3.8e-12& 2\\
Herschel:PACS70& 70.0    & 2.0e-11&3.9e-12& 2.0e-11& 3.9e-12& 3\\
Herschel:PACS100& 100.0 & 1.8e-11&3.0e-12& 1.8e-11& 3.0e-12& 3 \\
Herschel:PACS160 & 160.0   & 8.8e-12&1.7e-12& 8.8e-12& 1.7e-12& 3\\
LABOCA& 870.0 & 3.7e-13&3.4e-15& 3.7e-13& 3.4e-15& 4 \\ 
ALMA& 887.0 & 3.6e-13&3.4e-15& 3.6e-13& 3.4e-15& 5 \\
ALMA& 1300.0 & 1.0e-14& 1.0e-15& 1.0e-14& 1.0e-15& 6 \\
\enddata
\tablenotetext{}{\textbf{Note}—$\lambda F_\lambda$ in units of ergs cm$^{-2}$ s$^{-1}$. Columns 5 \& 6 are corrected for extinction. * indicates that the observation was dereddened using empirical extinction curves from \citet{mcclure_observational_2009}. All other observations are dereddened using CCM89. References: (1) \citet{fedorov_p_vizier_2011}; (2) \citet{luhman_disk_2008}; (3) \citet{ribas_2017}; (4) \citet{belloche_end_2011}; (5) \citet{pascucci_steeper_2016}; (6) \citet{hsieh_campos_2025}.}
\end{deluxetable*}

\section {Models} \label{models}
In this section we describe the models used to self-consistently 
obtain the mass accretion rate, the Ca abundance in the accretion flows, and the disk dust properties and structure.

\subsection{Accretion Shock Models}\label{sec:acc-shock-models}
We obtained the accretion luminosity and the mass accretion rate onto the star using an accretion shock model \citep[][CG98]{calvet_structure_1998}
to fit the excess flux over the photosphere in the visible range.
In this model, the inner disk is truncated by the 
stellar magnetic field at a radius $R_i$, and columns of accreting material travel along the magnetic field lines at the free fall velocity,
creating a shock when they hit the photosphere. 
The shock emits soft X-rays into the pre-shock and post-shock regions, as well as the photosphere below the shock. These regions get heated and produce the observed excess emission in the continuum. Following CG98,
the emission from the accretion column is characterized by the energy flux, $\curf$,  and the filling factor $f$. The energy flux is defined as $\curf = 1/2 \, \rho \, v_s^3$ 
where $\rho$ is the density of the pre-shock region, and $v_s$ 
is the free-fall velocity, defined as
\begin{equation}
     v_s = \left( \frac{2GM_*}{R_*} \right)^{1/2} \left( 1-\frac{R_*}{R_i} \right)^{1/2}
\end{equation}
where $M_*$ and $R_*$ are the stellar mass and radius.
The filling factor $f$ is the fraction of the stellar surface covered by the accretion column. As such, it is independent of wavelength and uniformly scales the luminosity of the shock emission. The mass accretion rate can be calculated 
as (CG98):
\begin{equation}
     \mathrm{\Mdot{}} = \frac{8 \pi R_*^2}{v_s^2} \curf f
\end{equation}

\subsection{Magnetospheric Accretion Models}
The magnetospheric accretion model is the current paradigm for accretion in TTS \citep[see][for a review]{hartmann_accretion_2016}. In this work, we use the {\it CV-multi} code \citep{Muzerolle_2001} to calculate the structure and emission of the magnetospheric accretion flows, which has successfully reproduced fluxes and profiles of the Balmer lines \citep{Muzerolle_2001, Thanathibodee_2023,Patino_2026} and the Ca II K line while also estimating the Ca abundance in the accretion flows \citep{micolta_ca_2023,Micolta_2024,micolta_vanishing_2025}.

The {\it CV-multi} models assume magnetospheric accretion flows that follow the stellar magnetic field lines, with dipolar geometry characterized by the inner radius ($R_i$) and the width at the base of the flow ($\rm W_r$) in the disk. The density profile is set by the system’s geometry and the mass accretion rate (\mdot). The temperature profile is semiempirical: a volumetric heating rate, scaling as $r^{-3}$, is equated to an optically thin radiative cooling law \citep[][]{1994ApJ...426..669H}. The temperature profile is parameterized in terms of the maximum flow temperature, \tmax{} which is constrained by the accretion rate according to the prescription of \citet[][]{Muzerolle_2001}.

We adopt the extended Sobolev approximation, which allows photons along the line of sight to interact with material on all the surfaces with the same velocity as that of the point where the photon was emitted, following \citet{1978ApJ...219..654R}, as implemented in \citet{1994ApJ...426..669H}. The models apply this approximation in order to compute the mean intensities, which are then used to determine the radiative rates in the statistical equilibrium equations. The H and Ca II populations are calculated by solving the statistical equilibrium equations for a 16-level H atom and of a 5-level Ca II atom, respectively, at each point in the magnetosphere, assuming all Ca is singly-ionized. The radiative rates are calculated using the extended Sobolev approximation for each two levels, using the populations from the previous iteration to estimate the optical depths; the procedure is iterated until convergence in the populations is achieved \citep{Muzerolle_2001, micolta_ca_2023}. Line source functions are calculated from the populations and line fluxes come from a ray-by-ray solution of the transfer equation, using resonance, Stark, and Van de Waals broadening \citep{Muzerolle_2001} with damping constants from \citet{1976ApJS...30....1V}.


\subsection{Disk models} \label{diskmodels}

We use the suite of D’Alessio Irradiated Accretion Disk 
\citep[DIAD,][]{dalessio_effects_2006} models to calculate the disk structure and fit the spectral energy distribution of T28. The DIAD model assumes a steady disk that is heated by viscous dissipation and stellar and accretion shock irradiation. Using these heating mechanisms, the radial and vertical density and temperature structures of the disk are self-consistently calculated from hydrostatic equilibrium. 
The viscosity of the disk is characterized by the dimensionless $\alpha$ parameter, assumed to be constant throughout the disk \citep{1973A&A....24..337S, dalessio_effects_2006}. We additionally assume that the gas and dust temperatures are equal, and solve the equations of energy conservation and transport \citep{dalessio_effects_2006}. 

We model the dust in the disk as two populations: Each population consists of grains of size $a$ where the distribution is given by $n(a) \propto a^{-3.5}$ between a minimum size $a_{min}=0.005$ $\mu$m and a maximum size $a_{max}$, which is a free parameter. The two populations have different vertical distributions: the large grains concentrate on the midplane, whereas the small grains dominate the upper layers of the disk. The dust distribution is characterized by the dimensionless dust settling parameter $\epsilon$, defined as the ratio between the dust-to-gas mass of small grains and the total dust-to-gas mass ratio, and depends on the abundances of the included species. Changing $\epsilon$ alters the total mass of small grains in the upper layers, and the mass of large grains in the midplane is adjusted accordingly such that the total dust-to-gas mass ratio is kept constant. 


The opacity depends on the chemical composition of dust grains. We include amorphous and crystalline silicates in our modeling. More specifically, we include olivine ($\text{Mg}_{2x}\text{Fe}_{2(1-x)}\text{SiO}_4$) and pyroxene ($\text{Mg}_{x}\text{Fe}_{(1-x)}\text{SiO}_3$) as well as their respective crystalline forms, forsterite and enstatite where the magnesium content is $x = \text{Mg / (Mg + Fe)}$. We calculate opacities using OPTool \citep{2021ascl.soft04010D}.
We use MIE theory for amorphous materials and use a continuous distribution of ellipsoids for crystalline materials. For olivine, the available Mg fractions are 40\% and 50\% while for pyroxene, the available Mg fractions are 40\%, 50\%, 60\%, 70\%, 80\%, 95\%, and 100\%. For forsterite, the available Mg fractions are 0\%, 95\%, and 100\%. Only one Mg abundance, 96\%, is available for enstatite. We use OpTool CHON amorphous material to calculate the mass absorption coefficients of organic material. We further include silica and water ice with opacities from \citet{2009ApJS..182..477S} and \citet{2001ApJ...553..321D} respectively.

The DIAD model calculates the emission from the disk between an inner and an outer radius. The inner radius generally corresponds to the dust destruction radius, defined as the point at which the temperature of optically thin dust is equal to the sublimation temperature of silicates. The emission from the frontally-illuminated edge of the disk at this radius (the ``Wall'') is calculated separately, assuming that it is optically thick and dust properties that could be different from the disk itself \citep{dalessio_models_2003}.

Disks with inner cleared out cavities (transitional disks) can be simulated with DIAD by starting the disk at a radius larger than the dust destruction radius, adjusted to fit the observations.
Similarly, DIAD can also be applied to disks with gaps (pre-transitional disks) by allowing for an optically thick inner disk inside the  of an outer disk.
For pre-transitional disks then two disk models are calculated, an ``inner disk'' and an ``outer disk'', each with its own wall at their inner edge, the ``inner wall'' and the ``outer wall''. The radius of the inner wall corresponds to the dust destruction radius \citep{2014prpl.conf..497E}.

Transitional and pre-transitional disk models often
require a region of  optically thin dust in the gap or cavity.
 This region is modeled by allowing dust to exist between some inner radius, $R_{min}$,  and an outer radius,
 $R_{max}$, inside the cavity and scaling
 the emissivity at each radius and at each wavelength to 
 the optical depth at 10 $\mu$m, 
 which can be calculated for an adopted dust mixture.
 The optical depth
 is assumed to vary with radius as
 $\tau_{10} =\tau_{10}^i(R_{min}/R)^{\rm Power}$,
 where
 $\tau_{10}^i$
 is the optical depth at 10 $\mu$m at $R_{min}$.
Both $\tau_{10}^i$ and 
${\rm Power}$ are free parameters obtained from fitting the observations. 
%
The
temperature of the dust is calculated from radiative equilibrium as
\begin{equation}
\kappa_{\text{P}}^* T_*^4 = \kappa_{\text{P}} T^4
\end{equation}
where $\kappa_{\text{P}}^*$ and $\kappa_{\text{P}}$ are the Planck mean opacities at the stellar and local wavelength ranges respectively, and $T_*$ is the stellar temperature \citep{dalessio_effects_2006}. The opacities are assumed to be the same everywhere in the region.

\section{Analysis and results} \label{analysis}

\subsection{Extinction}

Optical extinction estimates for T28 vary substantially, with published $\text{A}_{\text{V}}$ values ranging from 2.8 to 4.8 \citep[e.g.][]{manara2017, luhman_disk_2008}. We adopt the J-band extinction value from \citet{luhman_disk_2008}, which is determined to be $\text{A}_{\text{J}} =1.35$ using the color excess between 0.6 and 0.9 $\mu$m in optical spectra and converting it to extinction as $\mathrm{A_J} =E(0.6-0.9)/0.95$. We convert this value to visual extinction using the extinction law from \citet{cardelli_relationship_1989} (hereafter CCM89) to obtain $\text{A}_{\text{V}} = 4.79$. We additionally validate this value using the total hydrogen column density of the region: We use the total hydrogen column density map from \citet{hayashi_gas_2019} to obtain $N_H=1.05 \times 10^{22} \ {\rm cm^{-2}}$ in the line of sight to T28. From this, we obtain $\text{A}_{\text{V}} = 4.8$ using the Galactic relation between $\text{A}_{\text{V}}$ and $N_H$ from \citet{guver_relation_2009}. This value is consistent with that obtained from \citet{luhman_disk_2008}, and as such, we adopt $\text{A}_{\text{V}} = 4.8$ as our extinction value.

We note that the hydrogen column density $N_H$ in \citet{hayashi_gas_2019} is integrated through the molecular cloud along the line of sight, representing an upper limit. This limit is still appropriate due to the position of T28 within the molecular cloud: \citet{2021ApJ...919...35Z} determine the distance to the three-dimensional skeleton of the molecular cloud to be 181 pc at the location of T28. The inner dense component and outer diffuse component of the molecular cloud are $\sigma_1 = 2.9$ pc and $\sigma_2 = 10.2$ pc respectively. At a distance of 190 pc, T28 is 9 pc behind the skeleton, and at 3.1 $\sigma$ from the inner dense component and 0.9 $\sigma$ of the outer diffuse component. Given these, we conclude that T28 is towards the far side of the molecular cloud complex, with greater than 99.9\% of the inner dense component and 83\% of the outer diffuse component of the molecular cloud being in front of it. As such, we find it appropriate to adopt the aforementioned $N_H$ value to validate the extinction value obtained by \citet{luhman_stellar_2007}.


We apply two different extinction laws to the data using our constrained optical extinction of $\text{A}_{\text{V}} = 4.8$. We use the extinction function from CCM89, which is applicable to the optical and ultraviolet regions, and the infrared (IR) region up to 5 $\mu$m. CCM89 parametrizes the extinction curve with the total-to-selective extinction $\mathrm{R_V}$. We use the average value in the diffuse interstellar medium (DISM), $\mathrm{R_V} = 3.1$, for our extinction corrections in the aforementioned wavelength regions. We correct the 5 - 20 $\mu$m IR region using the empirical extinction curves from \citet{mcclure_observational_2009}. While the wavelength dependence of extinction at IR is weaker compared to optical and ultraviolet, previous work \citep{2005ApJ...619..931I,flaherty_infrared_2007, chapman_mid-infrared_2008} has demonstrated that extinction towards molecular clouds such as Chameleon I is different from DISM extinction due to grain growth. In particular, grain growth in molecular clouds affects the continuum level of extinction and modifies the shape of silicate features in IR \citep{2007ApJ...666L..73C}. As such, we find the two-law approach that corrects for molecular cloud specific extinction appropriate.


\subsection{Stellar Parameters}\label{sec:stellar}









Our overall modeling of the accretion shock, line profiles, and SED requires accurate stellar parameters. 
Therefore, we apply the code \rotfit\ \citep[][]{Frasca2006, Frasca2015} to the continuum-normalized MIKE spectrum and measure the atmospheric parameters, \teff, \logg, and \feh, as well as the radial velocity (RV) and projected rotational velocity (\vsini). \rotfit\ searches within a grid of templates to find those that minimize the $\chi^{2}$ relative to the target spectrum across twenty-eight 100\,\AA-wide spectral segments spanning the range 4000--6800\,\AA. Emission lines and spectral regions strongly affected by telluric absorption are masked during the fitting process. 
The template grid is the same as that adopted by the OACT (Osservatorio Astrofisico di Catania) node within \textit{Gaia}-ESO survey \citep[GES; see e.g.,][]{Randich2022}. It is composed of high-resolution spectra of $\sim$ 400 slowly rotating stars (\vsini\,$\leq 3\,{\rm km\,s}^{-1}$) characterized by low activity levels and well-determined stellar parameters. These spectra are retrieved from the ELODIE archive \citep[R $\simeq$ 42\,000,][]{Moultaka2004} and downgraded to the MIKE resolution 
(R $\simeq$ 32\,500) 
via convolution with a Gaussian kernel.

\rotfit\ first determines the RV of the target star via cross-correlation with the template spectra. Each template is then Doppler shifted to the target rest frame and resampled onto the wavelength grid of the target spectrum.  Next, the templates are iteratively broadened by convolution with a rotational profile with increasing \vsini\ until the minimum $\chi^{2}$ is achieved. This procedure is repeated by introducing an artificial veiling, $r_{\lambda}$, to the template spectra according to:
\begin{equation}
     \frac{F_\lambda^r}{F_{\rm c}}=\frac{(F_\lambda/F_{\rm c})_{wtts}+r_{\lambda}}{1+r_{\lambda}},
\end{equation}

\noindent{where $F_{\rm c}$ represents the continuum flux. 
The optimal veiling for each of the 28 analyzed spectral segments is determined by minimizing the $\chi^{2}$ residuals. We find that the veiling values decrease with increasing wavelength, a behavior that is highly consistent with a dominant emission contribution from an accretion hot spot. 
To obtain more robust results, we average the veiling values across five consecutive spectral segments, reporting the final mean values at 4200, 4500, 5000, 5500, 6000, and 6500 \AA{}. The final stellar parameters, reported in Table~\ref{tab:params}, are derived as the weighted averages of the results of the best ten templates within each individual segment. These are weighted according to both the $\chi^2$ and the amount of spectral information contained in each wavelength range. The latter is quantified by the total line absorption, $f_i=\int(F_{\lambda}/F_{\rm c}-1)d\lambda$.
We determine the best fit spectral type to be M0\,V, corresponding to the template star GJ\,488.
This result is in excellent agreement with the previous classification by \citet{luhman_stellar_2007}, whereas it contrasts with the K5 spectral type reported by \citet{Frasca2015}. 
To investigate this discrepancy, we re-analyzed that specific GIRAFFE-HR15N spectrum, tailoring the ROTFIT configuration to its specific characteristics. From this new run, we derived a K7 spectral type, which is in much better agreement with our current determination based on the higher-resolution MIKE spectrum. Moreover, the newly derived best fitting veiling at 6500 {\AA} ($r_{6500}=0.7\pm0.2$) is in close agreement with the value obtained from the MIKE data.
}

We derive the stellar radius by scaling the MIKE spectra to the continuum of WTTS template TWA25 as:
\begin{equation}
    R_* =  R_{\mathrm{wtts}} 
    \left (
    \frac{d_*}{d_{\mathrm{wtts}}}
    \right )
    \left [
    \left (\frac{1}{1+r_{\lambda}}
    \right )
    \left (
    \frac{F_{\lambda, \mathrm{cal}}}{F_{\mathrm{wtts}}}
    \right )
    \right ]^{1/2}
\end{equation}
where $F_{\lambda, \mathrm{cal}}$ and $F_{\mathrm{wtts}}$ are the dereddened continuum fluxes of T28 and TWA25, 
$d_*$ and $d_{\mathrm{wtts}}$ are the distances to T28 and TWA25, and $R_{\mathrm{wtts}}$ is the 
radius of TWA25. We thus obtain a stellar radius of $R = 2.54 \pm 0.14\ \rsun$. 

We calculate the inclination of the central star as:
\begin{equation}
    \sin i_* = \frac{v\sin i_*}{v_{eq}}
\end{equation},
following the standard method \citep[][and references therein]{2020AJ....159...81M} where $v_{eq} = 2 \pi R_*/P$, with $P$ being the rotation period. Using our previously determined values for \vsini{} and $R_*$, and with $P = 7.174$ days from \citet{2022AJ....164..137K}, we obtain $i_* = 37^\circ$.
The stellar parameters are shown in Table~\ref{tab:params}. 

\subsection{Disk parameters \label{sec: disk param}}
We adopted the disk radius from \citet{2023ASPC..534..539M} as $\text{R}_{\text{d}}=33 \ \text{au}$. We note that this is consistent with the extent of the emission in the ALMA observation in \citet{hsieh_alma_2024}.  We calculate the inclination assuming azimuthal symmetry. We use $\text{R}_{\text{min}}=0.241$" and $\text{R}_{\text{maj}}=0.260$" values from \citet{hsieh_alma_2024}, where cos($i$) = $R_{min}$ / $R_{maj}$. We obtain cos($i$) = 0.93, corresponding to disk inclination $i=22 \pm 16^\circ$. The disk parameters are shown in Table~\ref{tab:params}. 

\begin{deluxetable}{lcc}[t!]
\tablecaption{Stellar, accretion, and disk parameters\label{tab:params}}
\tablehead{
\colhead{Property} & \colhead{Value} & \colhead{Reference} }
\startdata
\multicolumn{3}{c}{Stellar Parameters}\\
\hline
Distance (\textit{d}) & 190.2$\pm$0.6 pc & 1 \\
Mass (\textit{M}) & 0.48 $\msun$ & 2 \\
Radius (\textit{$R_*$}) & 2.54$\pm$0.14 $\rsun$ & This work \\
Extinction (\textit{\av}) & 4.8 & This work\\
SpT & M0 & This work \\
$\mathrm{T}_{\mathrm{eff}}$ & 3963\,$\pm$\,140\,K & This work \\
\logg\      &  3.31\,$\pm$\,0.05\,dex  &  This work \\
\feh\       &  $-0.04\,\pm$\,0.07\,dex &  This work \\
$v\sin i_*$\     &  10.8\,$\pm$\,1.1\,km\,
s$^{-1}$   &  This work \\
RV          &  16.64\,$\pm$\,0.43\,km\,s$^{-1}$ &  This work \\
$r_{4200}$   &  2.10\,$\pm$\,0.38   &  This work \\
$r_{4500}$   &  1.51\,$\pm$\,0.44   &  This work \\
$r_{5000}$   &  1.16\,$\pm$\,0.44   &  This work \\
$r_{5500}$   &  1.09\,$\pm$\,0.16   &  This work \\
$r_{6000}$   &  0.64\,$\pm$\,0.24   &  This work \\
$r_{6500}$   &  0.77\,$\pm$\,0.19   &  This work \\
$i_*$ & 37\,$\pm$\,4$^\circ$ & This work \\
\hline
\multicolumn{3}{c}{Accretion Parameters}\\
\hline
Energy flux ($\curf$) & $3.16 \times 10^{12}$  & This work \\
Filling factor ($f$) & 0.01  & This work \\
Accretion rate (\mdot{}) & $(6.90 \pm 1.65)\times 10^{-7}$ $\msun \text{yr}^{-1}$ & This work \\
\hline
\multicolumn{3}{c}{Disk Parameters}\\
\hline
Disk radius $R_d$ & 33 au & 3 \\
Inclination ($i_\mathrm{disk}$) & $22 \pm 16^\circ$ & This work \\
\hline
\enddata
\tablenotetext{}{(1) \citet{GaiaDR3}; (2) \citet{manoj_spitzer_2011}; (3) \citet{2023ASPC..534..539M}}
\end{deluxetable}



\subsection{Accretion parameters}\label{sec:accparam}
We applied the accretion shock model
\citep[see \S \ref{sec:acc-shock-models};][]{calvet_structure_1998} to data from MIKE spectrograph on Magellan/Clay telescope, which we scaled using SOAR photometry in $griz$. To do so, we first obtained synthetic photometric flux through each filter $i$ as:
\begin{equation}
     F_{\mathrm{syn},i} = \frac{\int F_{\lambda}(\lambda)T_i(\lambda)\ d \lambda}{\int T_i(\lambda)\ d \lambda}
\end{equation}
where $F_{\lambda}(\lambda)$ is the observed flux from MIKE instrument at a given wavelength in instrumental units and $T_i(\lambda)$ is the transmission curve of each filter. We then obtained photometric fluxes $F_{\mathrm{obs}, i}$ from the SOAR magnitudes and used them to convert synthetic fluxes from MIKE spectra to photometric fluxes. To do so, we calculated a calibration factor per filter as $C_i = F_{\mathrm{obs}, i} /F_{\mathrm{syn},i}$ and fit the $C_i$ we obtained with a function $C(\lambda)$, which is the wavelength-dependent calibration function. We then applied $C(\lambda)$ to the full MIKE spectrum as:
\begin{equation}
F_{\lambda, \mathrm{cal}} = C(\lambda)F_{\lambda}(\lambda)
\end{equation}
where $F_{\lambda, \mathrm{cal}}$ is the flux calibrated MIKE spectrum in physical units.

We estimated the continuum from dereddened calibrated MIKE spectrum using a binned-median approach where median fluxes within the bins (each spanning 150 \AA) are computed. We then uniformly sampled the continuum between 3300 and 9000 \AA{} for easier comparison between predicted and observed fluxes. To represent the photosphere of the star, we used the spectrum of the WTTS TWA25, which we scaled to the distance and the radius of T28. We used a WTTS instead of a main sequence star to take into account the UV excess due to the chromosphere, which can be significant \citep{Ingleby_2011}. We applied the same binned-median approach to sample the continuum of TWA25. 

\begin{figure*}
    \centering
    \includegraphics{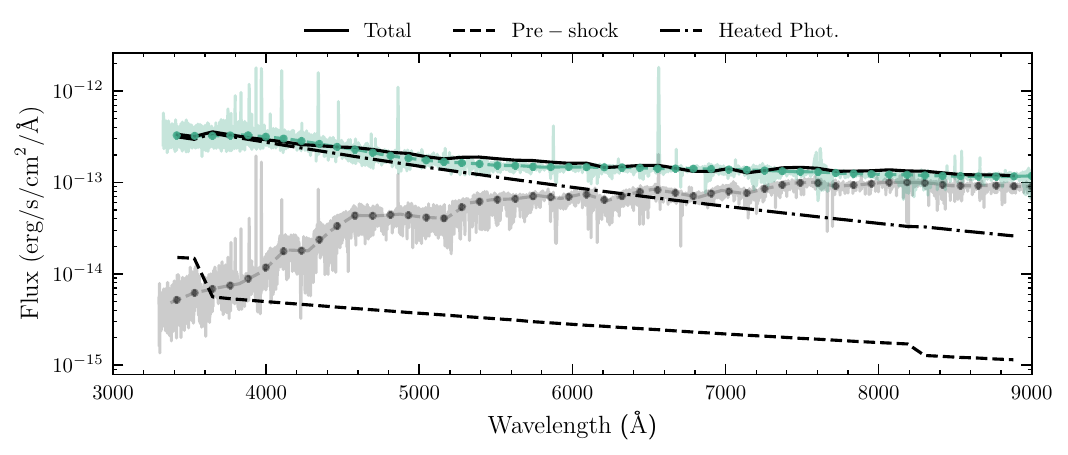}
    \caption{Accretion shock modeling of T28. The dereddened MIKE spectrum of T28 is shown in teal. The spectrum of the WTTS TWA25, used as the photospheric template, is shown in gray. The total emission output from the accretion shock model is shown as the black solid line. Emission from the pre-shock region (dashed line) and the heated photosphere (dash-dotted line) are also shown. The best fit parameters are $f=0.01$, $\curf=3.16 \times 10^{12}$, and \mdot{} $ = 6.9 \times 10^{-7}.$}
    \label{shock}
\end{figure*}

We used our mass and radius estimates for T28 (cf. Table \ref{tab:params}) to calculate a grid of shock models, covering the range log~$\curf$ from -10 to -12.5 in steps of 0.1 and log~$f$ = -3 to -1, in steps of 0.1 for a total of 546 models. 
We kept the disk truncation radius fixed at $R_i = 5 R_*$, following CG98. This value is consistent with the truncation radius inferred from fitting the line profiles (\S 4.5) within the uncertainties.
The total model emission is the sum of fluxes from the pre-shock and the heated photosphere, weighted by $f$, and the stellar photosphere, weighted by (1 - $f$). We used the $\chi^2$ statistics to find the best fit by comparing the emission from each model, binned in the same way as the observations, with the dereddened observed spectrum. The best fit model, shown in Figure \ref{shock}, yielded the value of \mdot{} as well as $\curf$ and $f$ shown in Table \ref{tab:params}. 
We estimated the uncertainty on \mdot{} to be $\pm1.65 \times 10^{-7}$, where we adopted a reduced-$\chi^2$ cutoff of $\chi^2_\nu \leq 3$ to select models with acceptable fits.

The best fit value for \mdot{} is
$6.90 \times 10^{-7} \msunyr$, a factor of $\sim$ 57 higher than the value of
$1.20 \times 10^{-8} \msunyr$ obtained by \citet{manara2017}. There are three factors that determine the increase: The most important one is the increase of A$_V$ by 2 magnitudes, which implies an increase in luminosity at 3300 {\AA} of $\sim 20$, using the standard reddening law, CCM89. In addition, we adopt a distance of 190 pc and obtain a radius of 2.54 $\rsun$, compared to 160 pc and 1.37 $\rsun $ adopted by \citet{manara2017}, resulting in an increase of $\sim 53$, sufficient to account for the difference.



\subsection{Ca abundance \label{sec:Ca-abun}}

\citet{micolta_ca_2023} identified T28 as a highly depleted star by analyzing the emission line fluxes of multiple Balmer and \CaII\ lines. However, this determination was done using the old literature parameters and magnetospheric accretion models made with stellar parameters of a generic M1, 3 Myr star. 
Here, we
revisit the Ca determination using the new MIKE observations, as well as the stellar and accretion parameters estimated in this work to generate models tailored to T28. Table \ref{tab:acc_model_param} shows the full parameter space explored. The Ca abundance values are cited relative to the solar value \citep[$\rm log \left(N_{Ca} / N_{H}\right)_{\odot} = 6.34 \pm 0.04$, ][]{2009ARA&A..47..481A}. The parameter space in abundance by number is $\rm \left(N_{Ca} / N_{H}\right) /\left(N_{Ca} / N_{H}\right)_{\odot} = $  1, 0.5,  0.1, 0.05, 0.01, 0.005, 0.001, 0.0005, 0.0001. The grid has a total of 142560 profiles.

To estimate magnetospheric accretion parameters and the Ca abundance, we perform detailed line profile modeling of the H$\alpha$ and Ca II K line. We follow the methodology of \citet{micolta_vanishing_2025}, and implement a maximum likelihood approach, introducing an error inflation parameter to account for the possible underestimation of observational uncertainties. The log-likelihood function, used to evaluate the fit of a model $F(x)$ to the observed data $y$, is given by:
\begin{equation}
\ln \mathcal{L}(y \mid x)=-\frac{1}{2} \sum_{i=1}^n\left[\frac{\left(y_i-F_i(x)\right)^2}{s_i^2}+\ln \left(2 \pi s_i^2\right)\right]
\label{eq:lk}
\end{equation}
where $n$ is the number of data points, $s_i^2$ is total variance at each data point $i$: $s_i^2=\sigma_i^2+10^b$, and $b$ is the error inflation parameter that reflects the degree of error adjustment necessary to explain the observed data.
The first term in Eq. \ref{eq:lk} is the $\chi^2$ that penalizes large residuals. The second term is the Gaussian normalization factor that accounts for the potential changes in the normalization factor due to error inflation and prevents the error inflation parameter from approaching infinity.

\begin{deluxetable}{lccc}[t!]
\tablecaption{Parameter space of the magnetospheric accretion models \label{tab:acc_model_param}}
\tablehead{
\colhead{Parameters} & \colhead{Min.} & \colhead{Max.} & \colhead{Step} 
}
\startdata
log {\mdot} ($\msunyr$)	    & -7.0 & -5.75 	& 0.25  \\
T$_{\rm max}$ (K)	        & 5000	& 10500	& 500 \\
R$_{\rm i}$	(R$_{\star}$)	& 1.5	& 5.5	& 1 \\
$\rm W_r$ (R$_{\star}$)	& 0.5   & 2.0   & 0.5 \\
$i_{\rm mag}$	(deg)				    & 15	& 65	& 5 \\ 
$[$Ca/H$]$$^*$	& -4	&  0	& ... \\ 
\hline
\enddata
\tablenotetext{*}{$\rm [Ca/H] = 
log_{10} (N_{Ca}/N_{H}) - log_{10} (N_{Ca}/N_{H})_{\sun}$.
Where 
$[$Ca/H$]=0$ is the solar abundance value.}
\end{deluxetable}


\begin{figure*}[t!]
\centering
\includegraphics[width = 0.85\textwidth]{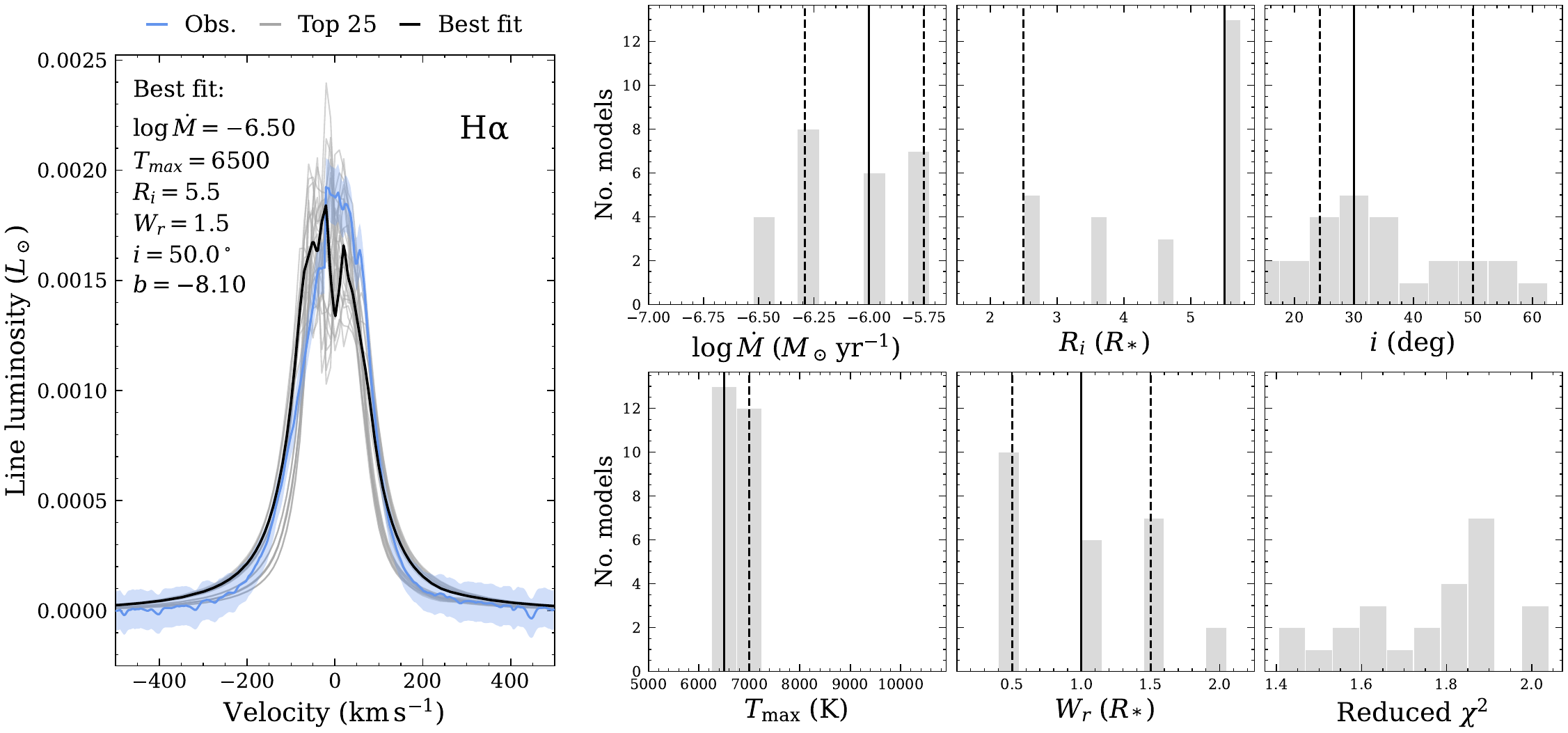}
\caption{H$\alpha$ profile modeling. Left: Comparison between the observations (blue), the best fit model (black), and the top-25 best-fitting models (gray). The blue shaded area is the flux uncertainty level. Right: Parameter distribution for the top 25 best-fitting models. The solid black line represents the median and the dashed lines represent the 16th and 84th percentiles.}\label{fig:h23-fit}
\end{figure*}

\begin{figure*}[t!]
\centering
\includegraphics[width = \textwidth]{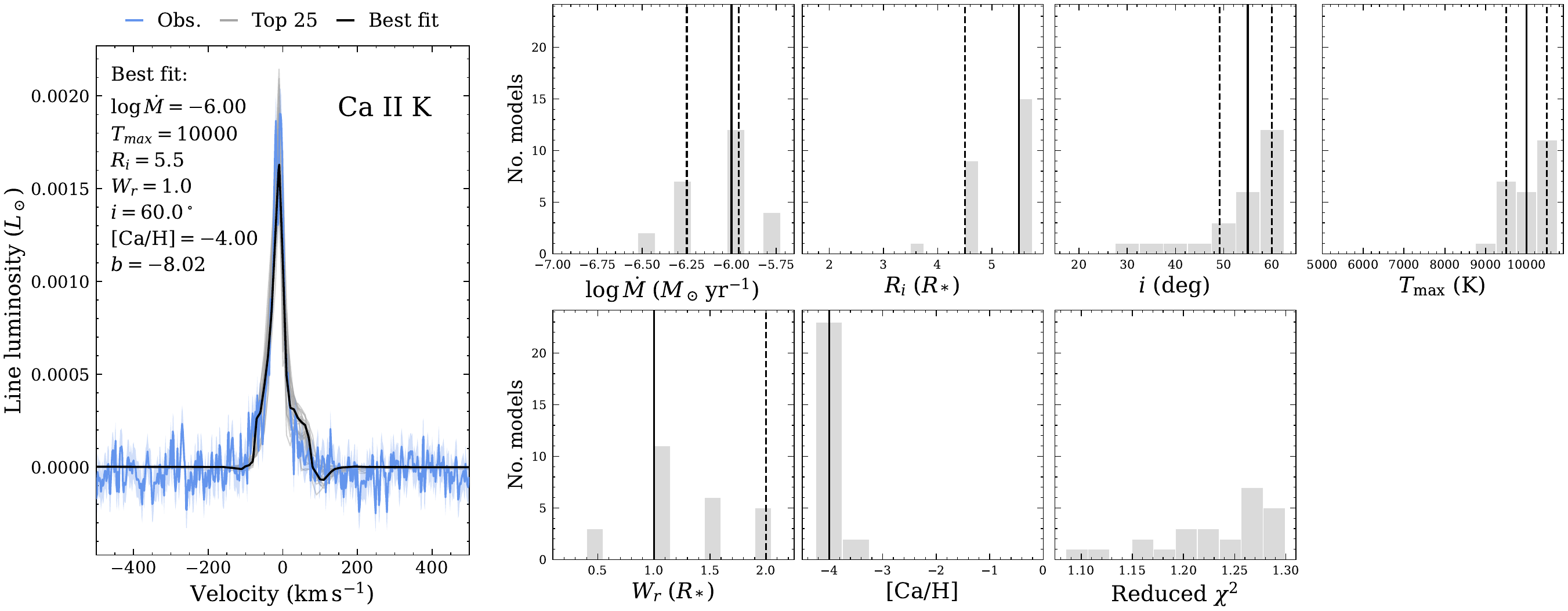}
\caption{\CaIIk\ profile modeling. Left: Comparison between the observations (blue), the best fit model (black), and the top-25 best-fitting models (gray). The blue shaded area is the flux uncertainty level. Right: Parameter distribution for the top 25 best-fitting models. The solid black line represents the median and the dashed lines represent the 16th and 84th percentiles.}\label{fig:ca15-fit}
\end{figure*}

\begin{deluxetable*}{rcccccccc}[t!]
\tablecaption{Line profile modeling results\label{tab:lines_results}}
\tablehead{\colhead{} & \colhead{$\rm Log\ \dot{\text{M}}$ ($\msunyr$)} & \colhead{T$_{\rm max}$ (K)} & \colhead{R$_{\rm i}$ (R$_{*}$)} & \colhead{W$_{\rm r}$ (R$_{*}$)} & \colhead{$i_{\rm mag}$ (deg)} & \colhead{ $\rm [Ca/H]$} & \colhead{ $\rm b$} & \colhead{ $\rm \chi^2_{\nu}$}}
\startdata
    \multicolumn{9}{c}{\halpha}\\
    \hline
    Best fit & $-6.50$  & 6500 & 5.5 & 1.5 & 50 & ... & $-8.10$ & 1.40\\
    Top 25  & $-6.00^{+0.25}_{-0.29}$  & 6500$^{+500}_{-0}$  & 5.50$^{+0}_{-3.00}$ & 1.00$^{+0.50}_{-0.50}$  & 30$^{+20}_{-5.8}$ & ... & $-8.10$ & $1.40-2.04$ \\
    \hline
    \multicolumn{9}{c}{\CaIIk}\\
    \hline
    Best fit&  $-6.0$ & 10000  & 5.5  & 1.0  & 60 &  $-4.0$ & $-8.02$ & 1.08\\
    Top 25  & $-6.0^{+0.05}_{-0.25}$  & 10000$^{+500}_{-500}$  & 5.50$^{+0.00}_{-1.00}$  &  1.0$^{+1.00}_{-0.00}$ & 55$^{+5}_{-5.8}$  & $-4.0^{+0.0}_{-0.0}$ & $-8.02$ & $1.08 - 1.30$\\
\enddata
\end{deluxetable*}

Figures \ref{fig:h23-fit} and \ref{fig:ca15-fit} show our line modeling results for H$\alpha$ and \CaIIk. In each figure, the left panel shows the observations (blue) as well as the top 25 best fitting models ranked according to the log likelihood (Eq. \ref{eq:lk}) in gray and the best fit model in black. From the top-25 best-fitting models, we calculate the median for the parameters and take the 16th and 84th percentiles as the uncertainties. The right-hand panels show the parameter distributions for the top 25 best-fitting models as well as their $\chi^2$.

In both cases, we can recover the observed line luminosities with the parameter values and errors shown in Table \ref{tab:lines_results}. We find that the mass accretion rates estimated from line profile modeling ($\rm log\ \dot{\text{M}} = 6.00^{+ 0.25}_{-0.29}$ for $\mathrm{H{\alpha}}$; $\rm log\ \dot{\text{M}} = 6.00^{+0.05}_{-0.25}$ for \CaIIk{}) are consistent with each other and with that estimated from the shock model $\rm log\ \dot{\text{M}} = 6.16 \pm 0.1$ within the uncertainties.

The best-fit model for H$\alpha$ has an inclination of $50^{\circ}$, which is higher than the one obtained from ALMA imaging ($22^{\circ}$, see \S \ref{sec: disk param}). However, the inclination distribution of the top-25 H$\alpha$ models yields inclinations of 30$^{+20}_{-5.8}$, which are consistent with the disk inclination results within errors. Additionally, these inclinations are in good agreement with the central star inclination derived in this work (see \S \ref{sec:stellar}). The differences with the disk inclination from ALMA can be explained considering that the magnetospheric accretion model assumes the magnetic axis is aligned with the rotation axis (of star and disk), when in reality these axes can be misaligned up to $20^{\circ}$ in TTS \citep{10.1111/j.1365-2966.2007.12194.x, 10.1111/j.1365-2966.2008.13111.x, 10.1111/j.1365-2966.2010.17409.x, 10.1111/j.1365-2966.2011.19288.x}, making the inclinations obtained by modeling magnetospheric models different from the ones obtained from disk imaging.

We find that the magnetospheric region is extended, with $\rm R_i \sim 5.5  R_*$ for the top models (Table \ref{tab:lines_results}).
Matter cannot fall onto the star along magnetic field lines that cut the disk beyond the corotation radius \citep{2018NewA...62...94R}, 
at which the disk Keplerian angular velocity equals the stellar angular velocity, given by
\begin{equation}
   R_c = \left(\frac{G M P^2}{4\pi^2}\right)^{1/3}
\end{equation}
 Using parameters given in section 4.2, $\rm R_c = 4.9 R_*$ for T28, so we do not explore values of $\rm R_i$ significantly beyond that point.
 The infalling region here is more extended than in the less detailed study by \citet{micolta_ca_2023}, which used generic stellar parameters, a lower value of the mass accretion accretion rate, and an inclination of 60$^o$.
However, the result that T28 is depleted in Ca holds, with the estimated abundance [Ca/H] dropping from $-0.76$ \citep{micolta_ca_2023} to $\rm [Ca/H] \leq -4.0$, indicating significant Ca depletion in the accretion flows. We report an upper limit for the Ca abundance, corresponding to the lower limit to the parameter space we explored. Similarly, the best values  of  T$_{\rm max}$ for the \CaIIk\ line are at the upper limit of the parameter space. In contrast, the values of the maximum temperature determined by the fit to H$\alpha$, T$_{\rm max} \sim $ 7000K, is consistent with the T$_{\rm max}$ vs. \mdot{} relation in \citet{Muzerolle_2001}. Discrepancies between the magnetospheric geometrical parameters obtained by fitting different lines have been reported before \citep{Patino_2026}, and interpreted as the result of trying to fit the emission of lines that form at different temperatures and densities using a simple, axially symmetric, dipolar geometry, while the actual magnetospheric flows are rather complex \citep{2024MNRAS.528.2883Z}.


\subsection{Modeling of the Spectral Energy Distribution}

\begin{figure*}[t]
    \centering

    \begin{subfigure}[t]{0.48\linewidth}
        \centering
        \includegraphics[width=\linewidth]{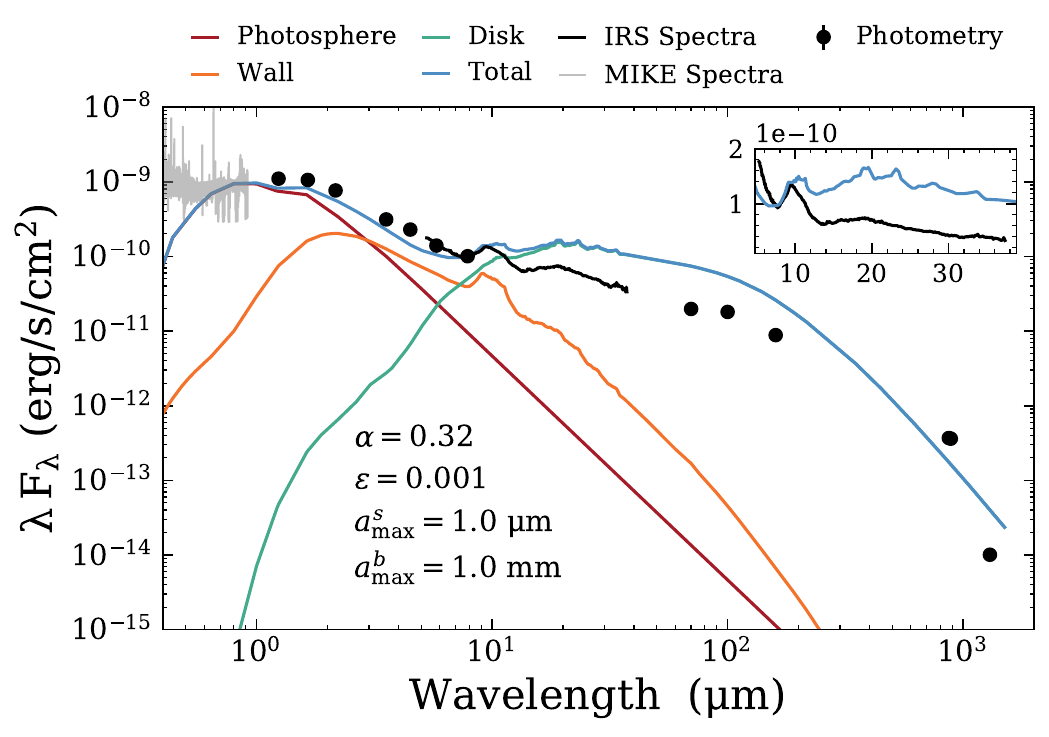}
    \end{subfigure}\hfill%
    \begin{subfigure}[t]{0.48\linewidth}
        \centering
        \includegraphics[width=\linewidth]{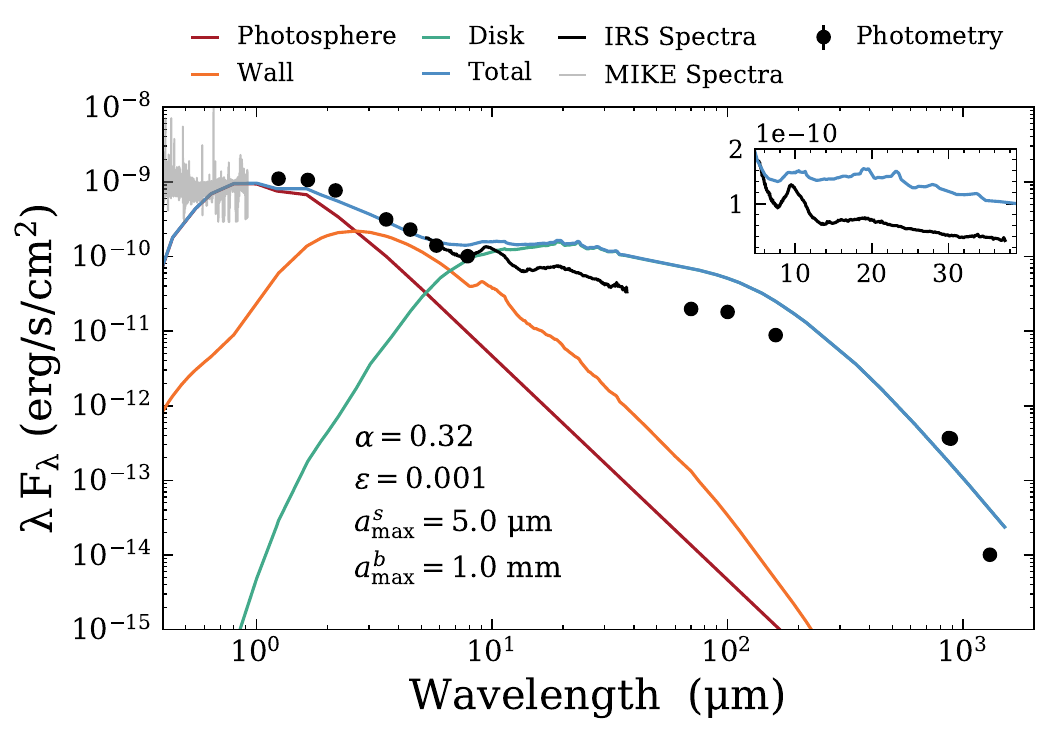}
    \end{subfigure}
    \caption{The spectral energy distribution of T28 alongside a two "typical" full-disk models with \mdot{} $= 5.0 \times 10^{-7} \msun{}\rm yr^{-1}$ and $i_{\rm disk} = 40^\circ$. The models differ in their size of the small grains population in the disk upper levels, where $a_{\text{max}}^s = 1.0 \ \mu$m for the left plot and $a_{\text{max}}^s = 5.0 \ \mu$m for the right plot. The silicate features are weaker in the model with the larger $a_{\text{max}}^s$.}
    \label{fig:bestfull}
\end{figure*}

The SED of T28 shows a significant excess over the photospheric fluxes, shown for reference, from the near-IR to the submillimeter region (Fig.~\ref{fig:bestfull}. In addition, the silicate features at 10 $\mu$m and 20 $\mu$m are conspicuous in emission in the IRS spectrum. We applied the DIAD models to find the disk parameters that best fit the SED of T28. We lack imaging data with high spatial resolution, which could inform usnof the presence of structure in the disk. Here, we discuss the viability of full disk and pre-transitional disk models in fitting the SED of T28.


We first attempted to apply a full disk model to explain the SED of T28. As described in section 3.3, the disk emission consists of the emission of the wall at the dust destruction radius and the disk proper. We adopted the mass accretion rate \mdot{} $= (6.90 \pm 1.65) \times 10^{-7} \msun{}\rm yr^{-1}$, obtained from fitting the flux excess with accretion shock emission and its uncertainty ($\S$\ref{sec:accparam} and Table \ref{tab:params}) and the disk inclination $i_{\rm disk} = 22 \pm 16^\circ$ provided by the ALMA data and its uncertainty ($\S$\ref{sec: disk param} and Table \ref{tab:params}). We varied the rest of the parameters, namely, the settling parameter $\epsilon$, the viscosity parameter $\alpha$, the maximum size of the populations of small and large grains, populating the disk upper levels and the midplane, respectively ($\S$\ref{diskmodels}).

The grain size of the small grain population in the disk upper levels is constrained by the strong silicate features, which only appear for $a_{max} \le 5 \mu$m (D’Alessio et al. 2006; Furlan et al. 2006).  This dust population controls the absorption of photons from the star and accretion shock and therefore the heating of the disk (D’Alessio et al 2006). With this constraint to the fitting procedure, in addition to the constraints on mass accretion rate and inclination,  we found no acceptable fits to the SED with a full disk model.

\begin{figure*}
    \centering
    \includegraphics[width=2
\columnwidth]{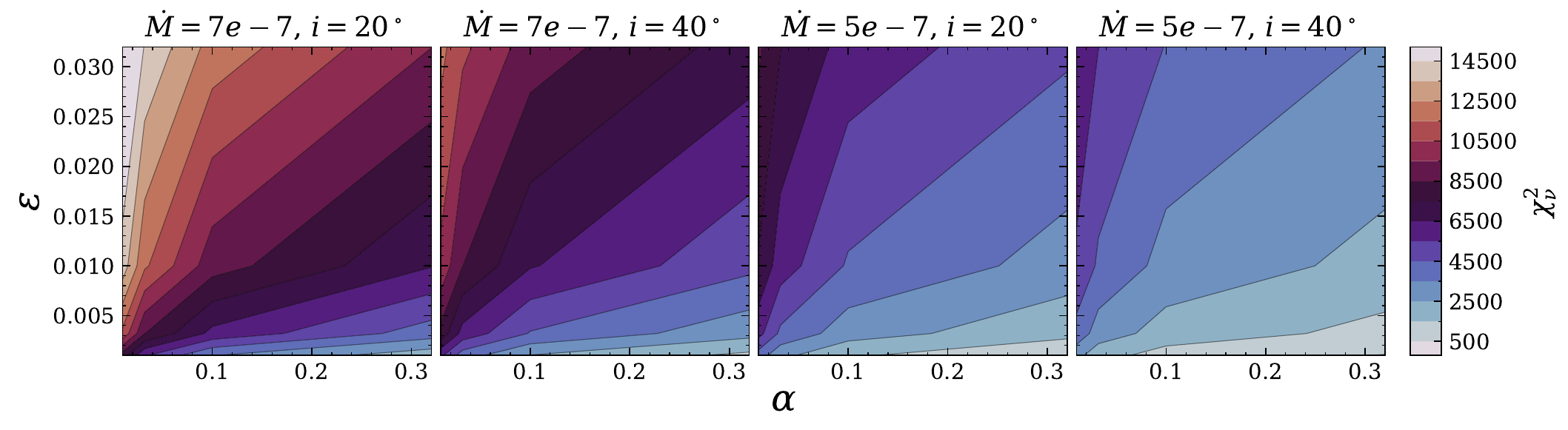}
    \caption{The values of the reduced chi-squared, $\chi^2_{\nu}$, as a function of $\alpha$ and $\epsilon$, evaluated by comparing model SEDs with observations. From left to right, the panels show $(\dot{M}, i_{\mathrm{disk}})=(7\times10^{-7}, 20)$, $(7\times10^{-7},40)$, $(5\times10^{-7},20)$, and $(5\times10^{-7},40)$ with $\dot{M}$ in $M_\odot\,\mathrm{yr}^{-1}$ and $i_{\mathrm{disk}}$ in degrees.}
    \label{contours}
\end{figure*}

The high value of \mdot{} and the low value of the inclination result in mid–IR fluxes too high compared to observations, requiring values of these parameters that are not consistent with observations. Increasing the grain size in the upper layer would also decrease the mid-IR emission, but then the silicate features would disappear. As an example of our search of parameter space, Fig.~\ref{contours} shows the value of $\chi_{\nu}^2$ - obtained by comparing the  observed SED with the model predictions (photosphere + wall + disk) - for fixed $a_{max} = 1 \mu$m for the small grains to get silicate emission, but values of \mdot{} and inclination corresponding to the limits provided by their uncertainty.  The value of $\chi_{\nu}^2$ does not fall below 500. 

Emission in the silicate features higher than what full models can predict has been reported before in analysis of IRS spectra of some disks and has been attributed to the presence of a gap in the disk filled with small dust, responsible for the strong silicate features \citep{Furlan2009, manoj_spitzer_2011, mauco_herschel_2018}. This motivates us to explore disks with structure to explain the SED of T28.


\begin{deluxetable}{lccc}[t!]
\tablecaption{Parameter space of the full disk models \label{tab:fulldiskparam}}
\tablehead{
\colhead{Parameters} & \colhead{Min.} & \colhead{Max.} & \colhead{Step} 
}
\startdata
\mdot{} ($\msunyr$)	        & $5.0 \times 10^{-7}$ & $8.5 \times 10^{-7}$ & $0.5 \times 10^{-7}$  \\
i$_{\rm disk}$ ($^\circ$)   & 20	& 40     & 5   \\
log $\alpha$	            & -3    & -0.5   & 0.5     \\
log $\epsilon$              & -3    & -0.5   & 0.5   \\
$a_{\text{max}}^s$ ($\mu$m) & 1.0   & 5.0    & $\ldots$  \\  
$a_{\text{max}}^b$ (mm)     & 1     & 20     & 10  \\  
\hline
\enddata
\end{deluxetable}

\subsubsection{Pre-transitional disk}





\begin{figure*}[ht]
\centering
\includegraphics{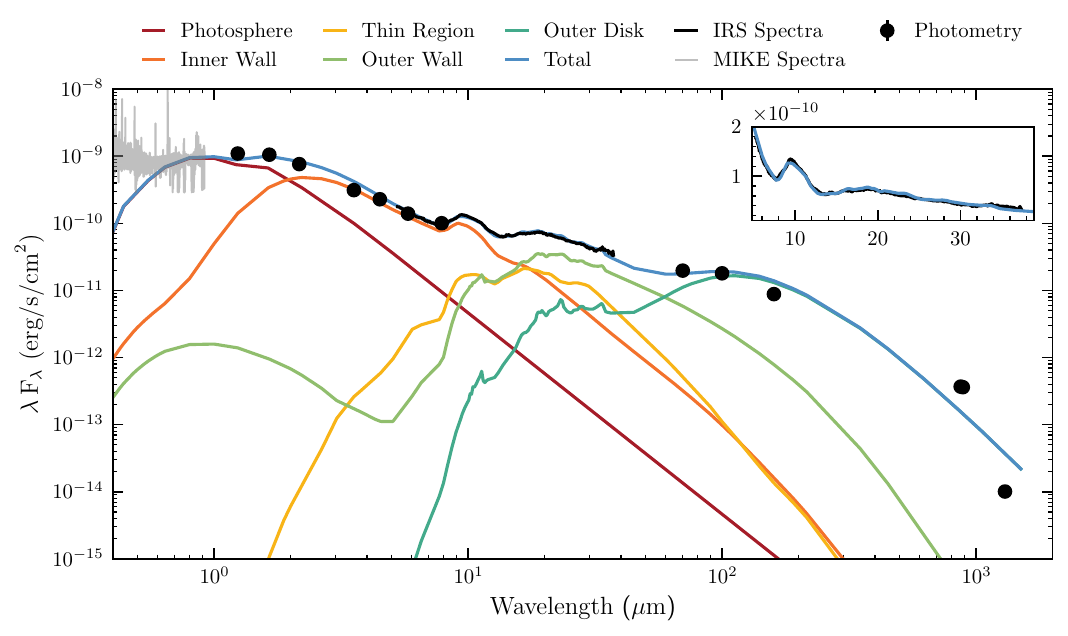}
\caption{SED of T28 with the best-fit pre-transitional disk model. The archival observations, consisting of photometry and Spitzer/IRS spectra, are shown in black. MIKE spectrum is shown in gray. The emission from the inner wall (orange), the optically thin region (yellow), the outer disk wall (green), and the outer disk (teal) are also shown. The photosphere is shown in red. The total emission from all components is shown as the blue line. The zoom-in at the top right corner shows the IRS spectra as well as the model fit.}\label{fig:diad}
\end{figure*}

Our model included two optically thick disks separated by a gap filled with optically thin dust (``Optically thin region''). We calculated a grid of models for the "Inner wall", representing the inner disk emission, and the outer disk, including its inner edge ("Outer wall") and the disk proper, and for the optically thin region. The grid that we used to determine the best fit consisted of 3456 inner wall, 864 outer wall, and 6912 optically thin region models. In all three grids, we keep the Mg fraction of olivine fixed to 50\%. The structural and dust parameter ranges, and the best fit parameters are listed in Table~\ref{tab:structure} and Table~\ref{tab:dust} The pre-transitional model SED is shown in Figure \ref{fig:diad}.

We determined the best fit in three steps: 
Most of the emission below 14 $\mu$m is due to the combined emissions from the photosphere, the inner wall, and the optically thin region. As such, we first fitted the inner wall and optically thin region components of our model to this region. 
Through this first step (hereafter, the inner model), and the photospheric contribution fixed, we obtained combinations of inner wall and optically thin models that could best recreate the emission below 14 $\mu$m. We then included the outer wall component in our fitting, and searched over top 100 best fit inner models up to 38 $\mu$m. Through this, we determined the best fit combination of the inner model and the outer wall. We finally added an outer disk component, keeping the previous three components fixed while assuming that the outer disk has the same composition as the outer wall. The resulting best fit model is shown in Fig.~\ref{fig:diad}. See \hyperref[appendix]{Appendix} for a more detailed discussion of the fitting routine.

\subsubsection{Inner Disk}
The inner disk emission is dominated by the frontally illuminated irradiated inner edge. We represent this as a vertical wall at the dust sublimation radius at which the temperature of optically thin dust is equal to the sublimation temperature of silicates, which we include as a free parameter and vary between 1300 - 1400 K. We set the maximum of this range such that it is below the sublimation temperature of the silicates we include in our model 
 \citep{lodders_solar_2003, kobayashi_sublimation_2011}. 
We find that the best fit grain size of the inner wall is 1 $\mu$m, corresponding to a dust sublimation radius of 0.4 au. The height of the wall $z_{wall, iw}$ was found to be 2.19 scale heights, or about 0.04  au.

We additionally obtain the stoichiometry of silicates in the inner wall by varying the fractional abundances of the constituents. We set our parameter space following the observed trends as per \citet{2009ApJS..182..477S} and \citet{2009ApJS..180...84W}. We assume a total  silicate-to-gas mass ratio of 0.004, and distribute it into olivine, pyroxene, forsterite, enstatite, and silica, in ranges indicated in Table \ref{tab:dust}.
We find that olivine makes up 95\% of the amorphous silicates, with the rest being made up of silica at 5\%. We find that the inner wall lacks forsterite, enstatite, and pyroxene.

\subsubsection{Outer Wall and Disk}
The outer wall defines the vertical and frontally illuminated inner edge of the outer disk. Similar to our treatments of the inner wall, we vary the temperature between 180 and 220 K and calculate the structure and emission of this region for a given stoichiometry. We find a best fit temperature of 190 K for the outer wall, corresponding to a radius of 16.9 au. In the gapped disk interpretation, we find the best fit for the height of the emitting region in the outer wall to be $z=0.17$ scale heights, corresponding to 0.36 au. This is smaller than the actual height of the disk at this radius, which is 3.17 au. This effect, commonly found in “pre-transitional” disks \citep{espaillat_unveiling_2010}, indicates that most of the 
region of the wall near the midplane
would be in the shadow created by the wall at the dust destruction radius 
and the emission would come just from a small region at the top of the outer disk edge.


At this wall temperature, the contribution to the opacity from water ice is negligible,
so we keep the H$_2$O ice mass ratio to gas fixed at 0.00001 and
explored a similar parameter space as in the inner wall (Table \ref{tab:dust}).
Unlike the inner wall, we find that pyroxene makes up the amorphous silicates in the outer wall, contributing 80\% to the total. Forsterite makes up the rest of the total at 20\%, with no enstatite  or olivine in the outer wall. We find the Mg fractions in pyroxene and forsterite to be 40\% and 100\% respectively.

We calculate the structure and emission of the outer disk with the stellar and accretion parameters shown in Table~\ref{tab:params}. We vary the dust settling $\epsilon$ and viscosity $\alpha$ parameters within the general range for CTTSs reported in \citet{rilinger_determining_2023} and obtain $\epsilon = 0.0002$ and $\alpha =0.1$. We explore the ranges of maximum sizes for the populations of small and large grains shown in Table \ref{tab:structure}, which also shows the adopted values. We adopt the same dust properties of the outer wall for the small grain population of the outer disk ($a_{\text{max}}^s$). 


\subsubsection{Optically thin region}

As in many pre-transitional disks \citep{espaillat_unveiling_2010}, we require a region of optically thin dust between the inner and outer disks. We model this region following the description in $\S$\ref{diskmodels}. We keep the minimum radius and the power
of the region fixed, and vary the outer radius and $\tau_{10}^i$ in the ranges shown in Table \ref{tab:structure}.


We find that the best-fit corresponds to $\tau_{10}^i = 0.01$
and $R_{max}=14$ au. We find that the composition of the optically thin dust is 20\% olivine and 80\% pyroxene. We determine the Mg abundances of pyroxene to be 40\%, same as the outer wall.

\begin{deluxetable}{lllr}[t!]
\tablecaption{Disk structural parameters \label{tab:structure}}
\tablehead{\colhead{Parameter} & \colhead{Range} & \colhead{Top 50} & \colhead{Best Fit}}
    \startdata
    \multicolumn{4}{c}{Inner Wall}\\
    \hline
    $T_{\text{wall,iw}}$ (K)    & 1300 - 1400 & $1375^{+25}_{-0}$    & 1375 \\
    $R_{\text{wall,iw}}$ (au)   & $\ldots$         & $\ldots$   $\ldots$& 0.4 \\
    $z_{\text{wall,iw}}/H$      & 0.5 - 4     & $2.20^{+0.08}_{-0.01}$   & 2.19 \\
    $z_{\text{wall,iw}}$ (au)   & $\ldots$         & $\ldots$   & 0.04     \\
    $a_{\text{max, iw}}$ ($\mu$m) & 0.25, 1   & $1^{+0.0}_{-0.0}$   & 1 \\
    \hline
    \multicolumn{4}{c}{Outer Wall}\\
    \hline
    $T_{\text{wall,ow}}$ (K)    & 180 - 220 & $190^{+10}_{-0.0}$      &  190 \\
    $R_{\text{wall,ow}}$ (au)   & $\ldots$       & $\ldots$     & 16.9 \\
    $z_{\text{wall,ow}}/H$      & 0.1 - 1   & $0.17^{+0.03}_{-0.03}$     & 0.17 \\
    $z_{\text{wall,ow}}$ (au)   & $\ldots$       & $\ldots$     & 0.36 \\
    $a_{\text{max,ow}}$ ($\mu$m) & 0.25, 1, 5  & $5^{+0.0}_{-0.0}$       & 5 \\
    \hline
    \multicolumn{4}{c}{Outer Disk}\\
    \hline
    $M_{\text{disk}}$ ($\msun{}$) & $\ldots$                  &  $\ldots$    & 0.12     \\
    $R_{\text{disk}}$ (au)      & $\ldots$             &  $\ldots$    & 33      \\
    $\alpha$                    & 0.01, 0.1, 0.15, 0.2 &  $\ldots$    & 0.1 \\
    $\epsilon$                  & 0.0002, 0.002, 0.02  &  $\ldots$    & 0.0002 \\
    $a_{\text{max}}^s$ ($\mu$m) & 0.25, 1, 5              &  $\ldots$    & 5    \\
    $a_{\text{max}}^b$ ($\mu$m) & 500, 1000, 10000, 20000 &  $\ldots$    & 10000    \\
    \hline
    \multicolumn{4}{c}{Optically Thin Region}\\
    \hline
    $R_{\text{min}}$ (au)       & $\ldots$                 &  $\ldots$      &  0.4     \\
    $R_{\text{max}}$ (au)       & 10, 12, 14, 16      &  $14^{+2}_{-0}$     &  14      \\
    Power                       & $\ldots$                &  $\ldots$       &  0.1     \\
    $\log \tau_{10}^i$          & $-2.3$, $-2$, $-1.3$, $-1$  &  $-2^{+0.0}_{-0.0}$   &    $-2$ \\
    $a_{\text{max,opt}}$ ($\mu$m) & 0.25, 1, 5 &  $5^{+0.0}_{-0.0}$    &  5 \\
    \enddata
\end{deluxetable}


\begin{deluxetable}{lllr}[t!]
\tablecaption{Dust properties \label{tab:dust}}
\tablehead{\colhead{Parameter} & \colhead{Range} & \colhead{Top 50} & \colhead{Best Fit}}
    \startdata
    \multicolumn{4}{c}{Inner Wall}\\
    \hline
    Organic/gas & 0.0025, 0.006 &  $0.006^{+0.0}_{-0.0}$      & 0.006\\
    H$_2$O      & ...           &   $\ldots$     & 0   \\
    Olivine     & 0 - 1         &  $0.950^{+0.0}_{-0.0}$ &  0.950  \\
    Pyroxene    & 0 - 1         &  $0^{+0.0}_{-0.0}$      &  0  \\
    Mg (pyr.)   & 40, 100       &  $\ldots$      &  $\ldots$   \\
    Forsterite  & 0 - 0.5       &  $0^{+0.0}_{-0.0}$     &  0 \\
    Mg (fors.)  & 0, 100        &  $\ldots$     &  $\ldots$  \\
    Enstatite   & 0 - 0.5       &  $0^{+0.0}_{-0.0}$     &  0 \\
    Silica      & 0, 0.05       &  $0.05^{+0.0}_{-0.0}$     &  0.05  \\
    \hline
    \multicolumn{4}{c}{Outer Wall}\\
    \hline
    Organic/gas     & $\ldots$        &     $\ldots$       & 0.0025        \\
    H$_2$O          & $\ldots$       &     $\ldots$  & 0.00001      \\
    Olivine         & 0 - 1     &   $0.0^{+0.16}_{-0.0}$      &  0   \\
    Pyroxene        & 0 - 1     &   $0.800^{+0.0}_{-0.16}$     &  0.800   \\
    Mg (pyr.)       & 40, 100   &   $40^{+9.6}_{-0}$              &  40   \\
    Forsterite      & 0 - 0.5   &   $0.200^{+0.0}_{-0.10}$    &  0.200  \\
    Mg (fors.)      & 0, 100    &   $100^{+0}_{-0}$             &  100   \\
    Enstatite       & 0 - 0.5   &   $0^{+0.10}_{-0.0}$           &  0  \\
    \hline
    \multicolumn{4}{c}{Optically Thin Region}\\
    \hline
    Organic/gas     & 0.0025, 0.006  &  $0.006^{+0.0}_{-0.0}$      & 0.006  \\
    Olivine         & 0 - 1          &  $0.200^{+0.0}_{-0.0}$        & 0.200      \\
    Pyroxene        & 0 - 1          &  $0.800^{+0.0}_{-0.0}$    & 0.800    \\
    Mg (pyr.)       & 40, 100        &  $40^{+0}_{-0}$             & 40       \\
    Forsterite      & 0 - 0.5        &  $0.00^{+0.0}_{-0.0}$      & 0    \\
    Mg (fors.)      & 0, 100         &  $\ldots$                  & $\ldots$ \\
    Enstatite       & 0 - 0.5        &  $0^{+0}_{-0}$              & 0      \\
    Silica          & 0, 0.05       &   $0.00^{+0.0}_{-0.0}$     & 0  \\
\enddata
\end{deluxetable}


\section{Discussion} \label{discussion}
T28 is the first high accretor that has been analyzed in the context of depletion and disk structure. Here we discuss accretion, structure, and depletion, and compare our findings to previous estimations. 

\subsection{Refractory depletion}
In \S\ref{sec:Ca-abun}, we showed that the Ca abundance in the magnetospheric flows of T28, and by extension, in its inner disk where the flows feed, is down by about four orders of magnitude relative to solar. Significant Ca depletion has been found in other low mass young objects in the Chamaeleon, Lupus, and Ori OB1b star-forming regions \citep{micolta_ca_2023,Micolta_2024,micolta_vanishing_2025}. This depletion is assumed to not have been inherited from the parent cloud (from which the disk and central star formed), but instead has been attributed to processes taking place in the disk: In a fully convective star such as T28, the surface abundance of the star is reflective of the initial abundance of the parent cloud \citep{2018MNRAS.476.4418J, 2018A&A...618A.132K}. It thus follows that if the surface abundance of the central star is solar, the composition of the parent cloud was originally solar.



Stellar surface abundances that are consistent with solar have previously been determined in  a small number of stars \citep{2008A&A...480..889S, 2011A&A...526A.103D}. In the case of T28, we  find \feh = $-0.04 \, \pm \, 0.07$ dex which is consistent with solar as well as the mean metallicity of the Chamaeleon I region \citep{Spina_2014}. Assuming that Fe is representative of other metals, the high level of depletion in Ca that we find in T28 cannot be explained by initial composition differences due to the parent cloud or the effects of stellar activity on emission lines \citep{Micolta_2024}. We thus attribute the depletion in the disk to dust-related processes.

\subsection{Refractory depletion and mass accretion rate}\label{sec:depletion-mdot}

\begin{figure}[t!]
\centering

\includegraphics[width=\columnwidth]{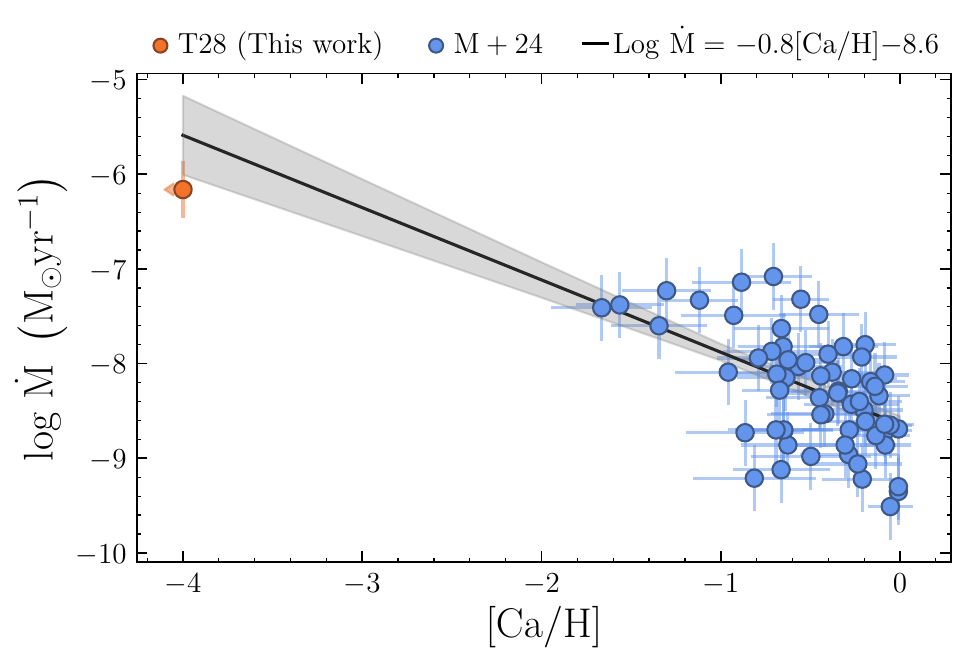}

\caption{$[\rm{Ca/H}]$ vs Log(\mdot). T28 is shown in orange. The values from the Cha I, Lupus and Orion OB1b stars (blue) are taken from \citet[][M+24]{Micolta_2024}. The black line represents the MCMC linear regression fitting. The gray region represents the $\rm 1\sigma$ confidence interval.}
\label{fig:mdot_cah}
\end{figure}

The depletion we determine for T28 is consistent with the trend of increased refractory depletion with higher mass accretion rate. Fig.~\ref{fig:mdot_cah} compares our result with previous determinations of Ca depletion of T Tauri stars in the Lupus, Chamaeleon, and Ori OB1b star forming regions \citep{micolta_vanishing_2025}. The reason for the increased refractory depletion in high mass accretion rate disks remains unknown. \citet{Micolta_2024} proposed that a dichotomy between the inner and outer disk could potentially be responsible, in that an inner disk mass reservoir could keep high accretion rates, even when the outer disk became highly depleted of refractories.  An alternative suggestion would be that pressure traps, holding
back refractory material in disks, are more efficient in disks with high accretion rates.  

Giant planets are required to open deep gaps in disks \citep{1986ApJ...309..846L,1997ApJ...482L.211W}, and to slow down
Type II migration of planets and gaps  towards the star.
\citep{2007prpl.conf..655P}.
In turn, high disk masses are among the most favorable conditions for giant planet formation \citep{2023A&A...679A..42S}. High mass accretion rate disks tend to have high disk masses \citep{2023ASPC..534..539M, 2025ApJ...989....5T}, suggesting that they would be more likely  to form giant planets and deep gaps. At the same time, gaps opened by giant planets trap pebbles \citep{lambrechts_rapid_2012, 2016A&A...585A..35P} and  the refractory material locked in them, so the presence of dust traps leads to increased refractory depletion in the material reaching the star, at different levels given the dependence of the trapping efficiency on planet mass and location, disk turbulence, and aspect ratio \citep{2018A&A...612A..30B}.

The correlation between higher depletion and both high {\mdot}  and possibly large dust masses \citep{Micolta_2024} could, in principle, arise because giant planets—and their associated pressure traps—are more likely to be present in disks with high mass-accretion rates, which also tend to have large disk masses.
To test the feasibility of this scenario, we need (1) more refractory abundance determinations in high mass accretion rate disks, to populate the high depletion end of the {\mdot} – [Ca/H] diagram (Fig.~\ref{fig:mdot_cah}), and (2) results from numerical simulations of refractory evolution in disks, spanning ranges of  stellar mass, viscosity parameter, initial disk sizes, and planet masses and locations consistent with planet demographics.


\subsection{Disk Structure and Dust Processing}

Our DIAD modeling results show that full disk models  
fail to reproduce the observed SED of T28. 
A pre-transitional disk model
\citep{2007ApJ...670L.135E} consisting of an optically thick
inner disk, to explain the near-IR excess, and an outer disk, to explain the far-IR to submillimeter fluxes, provides a much better fit to the observed SED. Optically thin dust filling the gap between the inner and outer disk is also required for the fit.

In the scenario presented in \S \ref{sec:depletion-mdot}, at least one giant planet would have been forming in the disk of T28, creating a gap at the edge of which refractory material would be trapped. In addition to the refractory material captured by the planet, this retention of rocky material in the disk would explain the depletion of Ca in the material reaching the star. The gap required by the SED fitting extends to 14 au, which at 190.2 pc requires at least an angular resolution $\theta = 0.074''$ to resolve. T28 has previously been observed with ALMA Band 6 as part of the CAMPOS large program which used a 0.1" synthesized beam, probing linear scales between 18 - 199 au across all disks \citep{hsieh_alma_2024}. Therefore, the cavity in the disk of T28 could not be resolved with ALMA, in general agreement with results of numerical simulations \citep{2025A&A...693A.302S}.

Our best-fitting models favor an outer disk with a particularly low $\varepsilon$ value, indicating that the dust in the disk is very settled, and the small-grain population in the upper layers is strongly depleted relative to the total dust content. It follows that there is a substantial degree of dust evolution in the disk, since grains have already grown and settled toward the midplane. Even with this degree of dust evolution, the mass inferred for the outer disk is high (Table \ref{tab:structure}), favoring the formation of giant planets \citep{2023A&A...679A..42S}.


Based on the detailed analysis of the infrared spectra of five T Tauri stars showing conspicuous features due to silica, 
\citet{2009ApJ...690.1193S} proposed that silica in protoplanetary disks is formed by 
melting of amorphous pyroxene
to olivine and a high temperature crystalline form of silica. In this context, it is interesting that in T28, pyroxene is found in the optically thin region, but only olivine and silica in the inner wall/disk. A process similar to that proposed by \citet{2009ApJS..182..477S} may be at play in the innermost disk of T28, which is expected to reach temperatures above 1000 K by viscous dissipation. In contrast, silica is not found in the inner wall/disk of BP Tau \citep{micolta_vanishing_2025}, which has a mass accretion rate about a factor of 20 lower than T28. Crystalline silica is found in cometary samples of the Stardust mission, with olivine generally associated with it \citep{2015ApJ...801L...7R}. However, these authors proposed an alternative path for the formation of silica rather than from pyroxenes. A detailed study of the possible pathways to form different mineral species in  protoplanetary disk environments is needed for a better understanding of the abundances, but it is beyond the scope of this paper.

\section{Summary and Conclusions} \label{summary}

In this work, we present a comprehensive study of the star T28 in the Chamaeleon I star-forming region. Previous studies had indicated that refractory depletion, using Ca as proxy,  was very high in this object,
and one of the goals of our work was to refine this determination. We adopted a higher extinction value than in previous studies, consistent with the star's location in the cloud, and carried out a detailed determination of the photospheric properties and of the mass accretion rate onto the star. 
We used magnetospheric models to estimate the Ca abundance relative to H in the magnetospheric flows, which feed from the inner disk.
Another goal of this work was to examine the dust composition in the disk surrounding the star. For this, 
we used the DIAD models to fit the SED and characterize the disk structure and dust properties, using the mass accretion rate and the inclination as constraints. 


Here we summarize the main results:

\begin{enumerate}

\item While the stellar abundance of T28 is consistent with solar and the Chameleon I region, there is considerable Ca depletion in the accretion flows.
We find an upper limit of -4 for [Ca/H].
%
    Therefore, the refractory depletion in the accretion flows, 
    cannot be due to initial composition. Instead, it is attributed to dust-related processes taking place in the disk.

   

    \item  We find an accretion rate about a factor of 50 higher than previous determinations. With this, T28 is the highest accretor for which refractory depletion has been determined so far, and confirms the trend of increasing depletion with increasing \mdot{}. One possible explanation for this trend is  preferential giant planet formation in massive disks leading to depletion. 
    This is related to \mdot{} through the observed correlation between \mdot{} and disk mass.
    
    \item The SED of T28 is best fit with a pre-transitional disk, instead of a full-disk model. Attempting to fit the SED with a full-disk model leads to IR excess that cannot be lowered without going beyond the error constraints of $i_{\mathrm{disk}}$ and \mdot{}. A better fit to the SED is obtained when a gap extending to 16 au, filled with optically thin dust up to 14 au. This gap would not be resolved in the existing observations.
    


    
    \item The best-fit DIAD model includes silica in the inner wall. It additionally includes olivine making up 95\% of the total dust composition in the inner wall at the dust destruction radius, as opposed to a lower olivine fraction, 20\% of total, and presence of pyroxene, 80\% of total, in the optically thin region filling the gap. This could potentially be explained by the high temperatures in the inner wall melting amorphous pyroxene to olivine and a crystalline form of silica. However, a detailed study of mineral formation pathways in protoplanetary disks is necessary to better understand the best-fit abundances.
\end{enumerate}

\begin{acknowledgments}
This research was supported in part by  NASA XRP 80NSSC2K0151. MM acknowledges funding from NASA FINESST 80NSSC24K1550. AF acknowledges funding from the Large Grant INAF-2024 ``Spectral Key features of Young stellar objects: Wind-Accretion LinKs Explored in the infraRed (SKYWALKER)''. E.M.M. acknowledges the support from a postdoctoral fellowship from the Secretaría de Ciencia, Tecnología, Humanidades e Innovación (SECIHTI), and the University of Michigan for the facilities provided for the development of this work.

Based in part on observations obtained at the Southern Astrophysical Research (SOAR) telescope, which is a joint project of the Minist\'{e}rio da Ci\^{e}ncia, Tecnologia e Inova\c{c}\~{o}es (MCTI/LNA) do Brasil, the US National Science Foundation’s NOIRLab, the University of North Carolina at Chapel Hill (UNC), and Michigan State University (MSU).

This research has made use of the NASA/IPAC Infrared Science Archive, which is funded by the National Aeronautics and Space Administration and operated by the California Institute of Technology.

\facilities{IRSA, Spitzer, Herschel, Magellan:Clay (MIKE spectrograph), SOAR (GHTS red camera), ALMA}

\end{acknowledgments}

\appendix \label{appendix}

\section{SED fitting algorithm}\label{sec:sedfitter}


In order to fit the spectral energy distribution, we use a fixed stellar photosphere and four dust-emission components, where each component has an associated amplitude. As such, the total model output is: 
\begin{equation}
  F_{\mathrm{model}}(\lambda)
  = F_\star(\lambda)
  + A_{\mathrm{iw}}\,F_{\mathrm{iw}}(\lambda)
  + A_{\mathrm{opt}}\,F_{\mathrm{opt}}(\lambda)
  + A_{\mathrm{ow}}\,F_{\mathrm{ow}}(\lambda)
  + \delta\,F_{\mathrm{od}}(\lambda),
  \label{eq:model}
\end{equation}
where $F_\star$ is the stellar photosphere and $F_{\mathrm{iw}}$, $F_{\mathrm{opt}}$,  $F_{\mathrm{ow}}$ and $F_{\mathrm{od}}$ are the inner-wall, optically-thin region, outer-wall, and outer disk templates respectively. Physically, $A_{\mathrm{iw}}$ and $A_{\mathrm{ow}}$ correspond to the wall height in units of the pressure scale height, $z_\mathrm{wall}/H$, and $A_{\mathrm{opt}}$ is the optical depth of the optically thin region at 10 $\mu$m, $\tau_{10}^i$. The parameter $\delta$ is a binary variable that determines whether we include the outer disk component in a given stage of the fitting routine. In other words, we do not fit to the amplitude of the outer disk. Instead, it is calculated self-consistently within the DIAD module.

The fitting problem thus consists of discrete and continuous parts: the choice of which four components to use is discrete, which we handle by searching over a precomputed grid of DIAD model outputs. The amplitudes are continuous, which we calculate exactly for each combination of components. The best fit model is therefore the minimum of $\chi^2$ over the [component, amplitude] space. The goodness-of-fit of a given combination of components is the weighted residual sum of squares:

\begin{equation}
  \chi^2 = \bigl\lVert \mathbf{b} - A_{\mathrm{iw}}\,\mathbf{w} - A_{\mathrm{opt}} \,\mathbf{o}
           - A_{\mathrm{ow}} \,\mathbf{u} - \delta \,\mathbf{v}\bigr\rVert^2 ,
  \label{eq:chi2}
\end{equation}
with the uncertainty ($\sigma$) weighted vectors
\begin{equation}
  \mathbf{b} = \frac{F_{\mathrm{obs}}-F_\star}{\sigma},\quad
  \mathbf{w} = \frac{F_{\mathrm{iw}}}{\sigma},\quad
  \mathbf{o} = \frac{F_{\mathrm{opt}}}{\sigma},\quad
  \mathbf{u} = \frac{F_{\mathrm{ow}}}{\sigma}, \quad
  \mathbf{v} = \frac{F_{\mathrm{od}}}{\sigma}.
  \label{eq:vectors}
\end{equation}
We minimize Eq.~\eqref{eq:chi2} over the amplitudes with the condition $A_{\mathrm{iw}}, A_{\mathrm{opt}}, A_{\mathrm{ow}} \ge 0$, as negative fluxes are nonphysical. Since Eq.~\eqref{eq:chi2} is a sum of squares, it has a single minimum, with a unique solution that is either at the unconstrained minimum or on the boundary where one amplitude vanishes. 

Between [$\lambda \in [1,14]\,\mu$m], the contributions of the outer wall and outer disk components are negligible. We thus only fit the inner wall and optically thin region components to this region, and set $A_{\mathrm{ow}}, \delta = 0$. For every inner wall model and optically thin region model pairing, we find the two amplitudes that minimize Eq.~\eqref{eq:chi2}, which has a closed-form solution. We keep the top 100 best pairs for the next stage of the fitting routine. We then include the outer wall, and extend the fit to 38 $\mu$m. We search over the top 100 best inner wall and optically thin model pairs by now including the outer wall component and solving for the three amplitudes, $A_{\mathrm{iw}}, A_{\mathrm{opt}}, A_{\mathrm{ow}}$. The outer disk component is excluded in this stage of the fitting routine ($\delta=0$). The three-component solution does not have a closed form. As such, we use a non-negative least squares (NNLS) routine. In particular, we use \texttt{scipy.optimize.nnls} and adopt the combination with the lowest $\chi^2$ as the best fit \citep{2020SciPy-NMeth}. 

Finally, we include the outer disk component ($\delta = 1$) and fit to the full wavelength range of the SED. We assume that the outer disk has the same composition as the best fit outer wall, and run a grid of models. We keep the best fit three-component combination from the previous stage of the fitting routine fixed, and solve for amplitudes $A_{\mathrm{iw}}, A_{\mathrm{opt}}, A_{\mathrm{ow}}$ again while searching over our grid of outer disk models using the aforementioned NNLS routine. The four-component solution with the lowest $\chi^2$ is adopted as the best fit. 



\bibliography{t28.bib}{}

@ARTICLE{2018NewA...62...94R,
       author = {{Romanova}, M.~M. and {Blinova}, A.~A. and {Ustyugova}, G.~V. and {Koldoba}, A.~V. and {Lovelace}, R.~V.~E.},
        title = "{Properties of strong and weak propellers from MHD simulations}",
      journal = {\na},
         year = 2018,
        month = jul,
       volume = {62},
        pages = {94-114},
          doi = {10.1016/j.newast.2018.01.011},
archivePrefix = {arXiv},
       eprint = {1704.08336},
 primaryClass = {astro-ph.SR},
       adsurl = {https://ui.adsabs.harvard.edu/abs/2018NewA...62...94R}
}

@ARTICLE{2024MNRAS.528.2883Z,
       author = {{Zhu}, Zhaohuan and {Stone}, James M. and {Calvet}, Nuria},
        title = "{A global 3D simulation of magnetospheric accretion - I. Magnetically disrupted discs and surface accretion}",
      journal = {\mnras},
         year = 2024,
        month = feb,
       volume = {528},
       number = {2},
        pages = {2883-2911},
          doi = {10.1093/mnras/stad3712},
archivePrefix = {arXiv},
       eprint = {2309.15318},
 primaryClass = {astro-ph.SR},
       adsurl = {https://ui.adsabs.harvard.edu/abs/2024MNRAS.528.2883Z}
}

@ARTICLE{2020AJ....159...81M,
       author = {{Masuda}, Kento and {Winn}, Joshua N.},
        title = "{On the Inference of a Star's Inclination Angle from its Rotation Velocity and Projected Rotation Velocity}",
      journal = {\aj},
         year = 2020,
        month = mar,
       volume = {159},
       number = {3},
          eid = {81},
        pages = {81},
          doi = {10.3847/1538-3881/ab65be},
archivePrefix = {arXiv},
       eprint = {2001.04973},
 primaryClass = {astro-ph.IM},
       adsurl = {https://ui.adsabs.harvard.edu/abs/2020AJ....159...81M}
}

@ARTICLE{Furlan2009,

       author = {{Furlan}, E. and {Watson}, Dan M. and {McClure}, M.~K. and {Manoj}, P. and {Espaillat}, C. and {D'Alessio}, P. and {Calvet}, N. and {Kim}, K.~H. and {Sargent}, B.~A. and {Forrest}, W.~J. and {Hartmann}, L.},

        title = "{Disk Evolution in the Three Nearby Star-forming Regions of Taurus, Chamaeleon, and Ophiuchus}",

      journal = {\apj},

         year = 2009,

        month = oct,

       volume = {703},

       number = {2},

        pages = {1964-1983},

          doi = {10.1088/0004-637X/703/2/1964},

archivePrefix = {arXiv},

       eprint = {0909.0960},

 primaryClass = {astro-ph.SR},

       adsurl = {https://ui.adsabs.harvard.edu/abs/2009ApJ...703.1964F}

}

@ARTICLE{2020ApJ...893...56M,
       author = {{Manzo-Mart{\'\i}nez}, Ezequiel and {Calvet}, Nuria and {Hern{\'a}ndez}, Jes{\'u}s and {Lizano}, Susana and {Hern{\'a}ndez}, Ramiro Franco and {Miller}, Christopher J. and {Mauc{\'o}}, Karina and {Brice{\~n}o}, C{\'e}sar and {D'Alessio}, Paola},
        title = "{The Evolution of the Inner Regions of Protoplanetary Disks}",
      journal = {\apj},
         year = 2020,
        month = apr,
       volume = {893},
       number = {1},
          eid = {56},
        pages = {56},
          doi = {10.3847/1538-4357/ab7ead},
archivePrefix = {arXiv},
       eprint = {2004.02916},
 primaryClass = {astro-ph.SR},
       adsurl = {https://ui.adsabs.harvard.edu/abs/2020ApJ...893...56M}
}

@ARTICLE{2006ApJS..165..568F,
       author = {{Furlan}, E. and {Hartmann}, L. and {Calvet}, N. and {D'Alessio}, P. and {Franco-Hern{\'a}ndez}, R. and {Forrest}, W.~J. and {Watson}, D.~M. and {Uchida}, K.~I. and {Sargent}, B. and {Green}, J.~D. and {Keller}, L.~D. and {Herter}, T.~L.},
        title = "{A Survey and Analysis of Spitzer Infrared Spectrograph Spectra of T Tauri Stars in Taurus}",
      journal = {\apjs},
         year = 2006,
        month = aug,
       volume = {165},
       number = {2},
        pages = {568-605},
          doi = {10.1086/505468},
archivePrefix = {arXiv},
       eprint = {astro-ph/0608038},
 primaryClass = {astro-ph},
       adsurl = {https://ui.adsabs.harvard.edu/abs/2006ApJS..165..568F}
}

@ARTICLE{2004A&A...421.1075D,
       author = {{Dullemond}, C.~P. and {Dominik}, C.},
        title = "{The effect of dust settling on the appearance  of protoplanetary disks}",
      journal = {\aap},
         year = 2004,
        month = jul,
       volume = {421},
        pages = {1075-1086},
          doi = {10.1051/0004-6361:20040284},
archivePrefix = {arXiv},
       eprint = {astro-ph/0405226},
 primaryClass = {astro-ph},
       adsurl = {https://ui.adsabs.harvard.edu/abs/2004A&A...421.1075D}
}

@ARTICLE{manara2017,
       author = {{Manara}, C.~F. and {Testi}, L. and {Herczeg}, G.~J. and {Pascucci}, I. and {Alcal{\'a}}, J.~M. and {Natta}, A. and {Antoniucci}, S. and {Fedele}, D. and {Mulders}, G.~D. and {Henning}, T. and {Mohanty}, S. and {Prusti}, T. and {Rigliaco}, E.},
        title = "{X-shooter study of accretion in Chamaeleon I. II. A steeper increase of accretion with stellar mass for very low-mass stars?}",
      journal = {\aap},
         year = 2017,
        month = aug,
       volume = {604},
          eid = {A127},
        pages = {A127},
          doi = {10.1051/0004-6361/201630147},
archivePrefix = {arXiv},
       eprint = {1704.02842},
 primaryClass = {astro-ph.SR},
       adsurl = {https://ui.adsabs.harvard.edu/abs/2017A&A...604A.127M}
}

@ARTICLE{1978ApJ...219..654R,
       author = {{Rybicki}, G.~B. and {Hummer}, D.~G.},
        title = "{A generalization of the Sobolev method for flows with nonlocal radiative coupling.}",
      journal = {\apj},
         year = 1978,
        month = jan,
       volume = {219},
        pages = {654-675},
          doi = {10.1086/155826},
       adsurl = {https://ui.adsabs.harvard.edu/abs/1978ApJ...219..654R}
}

@ARTICLE{1997ApJ...482L.211W,

       author = {{Ward}, William R.},

        title = "{Survival of Planetary Systems}",

      journal = {\apjl},

         year = 1997,

        month = jun,

       volume = {482},

       number = {2},

        pages = {L211-L214},

          doi = {10.1086/310701},

       adsurl = {https://ui.adsabs.harvard.edu/abs/1997ApJ...482L.211W}

}

@ARTICLE{1986ApJ...309..846L,

       author = {{Lin}, D.~N.~C. and {Papaloizou}, John},

        title = "{On the Tidal Interaction between Protoplanets and the Protoplanetary Disk. III. Orbital Migration of Protoplanets}",

      journal = {\apj},

         year = 1986,

        month = oct,

       volume = {309},

        pages = {846},

          doi = {10.1086/164653},

       adsurl = {https://ui.adsabs.harvard.edu/abs/1986ApJ...309..846L}

}

@INPROCEEDINGS{2007prpl.conf..655P,

       author = {{Papaloizou}, J.~C.~B. and {Nelson}, R.~P. and {Kley}, W. and {Masset}, F.~S. and {Artymowicz}, P.},

        title = "{Disk-Planet Interactions During Planet Formation}",

    booktitle = {Protostars and Planets V},

         year = 2007,

       editor = {{Reipurth}, Bo and {Jewitt}, David and {Keil}, Klaus},

        month = jan,

        pages = {655},

          doi = {10.48550/arXiv.astro-ph/0603196},

archivePrefix = {arXiv},

       eprint = {astro-ph/0603196},

 primaryClass = {astro-ph},

       adsurl = {https://ui.adsabs.harvard.edu/abs/2007prpl.conf..655P}

}

@ARTICLE{2016A&A...585A..35P,

       author = {{Pinilla}, P. and {Klarmann}, L. and {Birnstiel}, T. and {Benisty}, M. and {Dominik}, C. and {Dullemond}, C.~P.},

        title = "{A tunnel and a traffic jam: How transition disks maintain a detectable warm dust component despite the presence of a large planet-carved gap}",

      journal = {\aap},

         year = 2016,

        month = jan,

       volume = {585},

          eid = {A35},

        pages = {A35},

          doi = {10.1051/0004-6361/201527131},

archivePrefix = {arXiv},

       eprint = {1511.04105},

 primaryClass = {astro-ph.EP},

       adsurl = {https://ui.adsabs.harvard.edu/abs/2016A&A...585A..35P}

}

@ARTICLE{2018A&A...612A..30B,

       author = {{Bitsch}, Bertram and {Morbidelli}, Alessandro and {Johansen}, Anders and {Lega}, Elena and {Lambrechts}, Michiel and {Crida}, Aur{\'e}lien},

        title = "{Pebble-isolation mass: Scaling law and implications for the formation of super-Earths and gas giants}",

      journal = {\aap},

         year = 2018,

        month = apr,

       volume = {612},

          eid = {A30},

        pages = {A30},

          doi = {10.1051/0004-6361/201731931},

archivePrefix = {arXiv},

       eprint = {1801.02341},

 primaryClass = {astro-ph.EP},

       adsurl = {https://ui.adsabs.harvard.edu/abs/2018A&A...612A..30B}

}

@ARTICLE{2021ApJ...919...35Z,
       author = {{Zucker}, Catherine and {Goodman}, Alyssa and {Alves}, Jo{\~a}o and {Bialy}, Shmuel and {Koch}, Eric W. and {Speagle}, Joshua S. and {Foley}, Michael M. and {Finkbeiner}, Douglas and {Leike}, Reimar and {En{\ss}lin}, Torsten and {Peek}, Joshua E.~G. and {Edenhofer}, Gordian},
        title = "{On the Three-dimensional Structure of Local Molecular Clouds}",
      journal = {\apj},
         year = 2021,
        month = sep,
       volume = {919},
       number = {1},
          eid = {35},
        pages = {35},
          doi = {10.3847/1538-4357/ac1f96},
archivePrefix = {arXiv},
       eprint = {2109.09765},
 primaryClass = {astro-ph.GA},
       adsurl = {https://ui.adsabs.harvard.edu/abs/2021ApJ...919...35Z}
}

@ARTICLE{2022AJ....164..137K,
       author = {{Kounkel}, Marina and {Stassun}, Keivan G. and {Bouma}, Luke G. and {Covey}, Kevin and {Hillenbrand}, Lynne A. and {Curtis}, Jason Lee},
        title = "{Untangling the Galaxy. IV. Empirical Constraints on Angular Momentum Evolution and Gyrochronology for Young Stars in the Field}",
      journal = {\aj},
         year = 2022,
        month = oct,
       volume = {164},
       number = {4},
          eid = {137},
        pages = {137},
          doi = {10.3847/1538-3881/ac866d},
archivePrefix = {arXiv},
       eprint = {2206.13545},
 primaryClass = {astro-ph.SR},
       adsurl = {https://ui.adsabs.harvard.edu/abs/2022AJ....164..137K}
}

@ARTICLE{briceno2019,
       author = {{Brice{\~n}o}, C{\'e}sar and {Calvet}, Nuria and {Hern{\'a}ndez}, Jes{\'u}s and {Vivas}, A. Katherina and {Mateu}, Cecilia and {Downes}, Juan Jos{\'e} and {Loerincs}, Jaqueline and {P{\'e}rez-Blanco}, Alice and {Berlind}, Perry and {Espaillat}, Catherine and {Allen}, Lori and {Hartmann}, Lee and {Mateo}, Mario and {Bailey}, III, John I.},
        title = "{The CIDA Variability Survey of Orion OB1. II. Demographics of the Young, Low-mass Stellar Populations}",
      journal = {\aj},
         year = 2019,
        month = feb,
       volume = {157},
       number = {2},
          eid = {85},
        pages = {85},
          doi = {10.3847/1538-3881/aaf79b},
       adsurl = {https://ui.adsabs.harvard.edu/abs/2019AJ....157...85B}
}

@ARTICLE{2021A&A...654A..72S,
       author = {{Schneider}, Aaron David and {Bitsch}, Bertram},
        title = "{How drifting and evaporating pebbles shape giant planets. II. Volatiles and refractories in atmospheres}",
      journal = {\aap},
         year = 2021,
        month = oct,
       volume = {654},
          eid = {A72},
        pages = {A72},
          doi = {10.1051/0004-6361/202141096},
archivePrefix = {arXiv},
       eprint = {2109.03589},
 primaryClass = {astro-ph.EP},
       adsurl = {https://ui.adsabs.harvard.edu/abs/2021A&A...654A..72S}
}

@ARTICLE{2021A&A...654A..71S,
       author = {{Schneider}, Aaron David and {Bitsch}, Bertram},
        title = "{How drifting and evaporating pebbles shape giant planets. I. Heavy element content and atmospheric C/O}",
      journal = {\aap},
         year = 2021,
        month = oct,
       volume = {654},
          eid = {A71},
        pages = {A71},
          doi = {10.1051/0004-6361/202039640},
archivePrefix = {arXiv},
       eprint = {2105.13267},
 primaryClass = {astro-ph.EP},
       adsurl = {https://ui.adsabs.harvard.edu/abs/2021A&A...654A..71S}
}

@ARTICLE{2013A&A...551A.107M,
       author = {{Manara}, C.~F. and {Testi}, L. and {Rigliaco}, E. and {Alcal{\'a}}, J.~M. and {Natta}, A. and {Stelzer}, B. and {Biazzo}, K. and {Covino}, E. and {Covino}, S. and {Cupani}, G. and {D'Elia}, V. and {Randich}, S.},
        title = "{X-shooter spectroscopy of young stellar objects. II. Impact of chromospheric emission on accretion rate estimates}",
      journal = {\aap},
         year = 2013,
        month = mar,
       volume = {551},
          eid = {A107},
        pages = {A107},
          doi = {10.1051/0004-6361/201220921},
archivePrefix = {arXiv},
       eprint = {1301.3058},
 primaryClass = {astro-ph.GA},
       adsurl = {https://ui.adsabs.harvard.edu/abs/2013A&A...551A.107M}
}

@ARTICLE{2007ApJ...666L..73C,
       author = {{Chiar}, J.~E. and {Ennico}, K. and {Pendleton}, Y.~J. and {Boogert}, A.~C.~A. and {Greene}, T. and {Knez}, C. and {Lada}, C. and {Roellig}, T. and {Tielens}, A.~G.~G.~M. and {Werner}, M. and {Whittet}, D.~C.~B.},
        title = "{The Relationship between the Optical Depth of the 9.7 {\ensuremath{\mu}}m Silicate Absorption Feature and Infrared Differential Extinction in Dense Clouds}",
      journal = {\apjl},
         year = 2007,
        month = sep,
       volume = {666},
       number = {2},
        pages = {L73-L76},
          doi = {10.1086/521789},
archivePrefix = {arXiv},
       eprint = {0707.3480},
 primaryClass = {astro-ph},
       adsurl = {https://ui.adsabs.harvard.edu/abs/2007ApJ...666L..73C}
}

@ARTICLE{2005ApJ...619..931I,
       author = {{Indebetouw}, R. and {Mathis}, J.~S. and {Babler}, B.~L. and {Meade}, M.~R. and {Watson}, C. and {Whitney}, B.~A. and {Wolff}, M.~J. and {Wolfire}, M.~G. and {Cohen}, M. and {Bania}, T.~M. and {Benjamin}, R.~A. and {Clemens}, D.~P. and {Dickey}, J.~M. and {Jackson}, J.~M. and {Kobulnicky}, H.~A. and {Marston}, A.~P. and {Mercer}, E.~P. and {Stauffer}, J.~R. and {Stolovy}, S.~R. and {Churchwell}, E.},
        title = "{The Wavelength Dependence of Interstellar Extinction from 1.25 to 8.0 {\ensuremath{\mu}}m Using GLIMPSE Data}",
      journal = {\apj},
         year = 2005,
        month = feb,
       volume = {619},
       number = {2},
        pages = {931-938},
          doi = {10.1086/426679},
archivePrefix = {arXiv},
       eprint = {astro-ph/0406403},
 primaryClass = {astro-ph},
       adsurl = {https://ui.adsabs.harvard.edu/abs/2005ApJ...619..931I}
}

@article{10.1111/j.1365-2966.2011.19288.x,
    author = {Donati, J.-F. and Gregory, S. G. and Alencar, S. H. P. and Bouvier, J. and Hussain, G. and Skelly, M. and Dougados, C. and Jardine, M. M. and M{\'{e}}nard, F. and Romanova, M. M. and Unruh, Y. C. and the MaPP collaboration},
    title = {The large-scale magnetic field and poleward mass accretion of the classical T Tauri star TW Hya},
    journal = {Monthly Notices of the Royal Astronomical Society},
    volume = {417},
    number = {1},
    pages = {472-487},
    year = {2011},
    month = {10},
    issn = {0035-8711},
    doi = {10.1111/j.1365-2966.2011.19288.x},
    url = {https://doi.org/10.1111/j.1365-2966.2011.19288.x},
    eprint = {https://academic.oup.com/mnras/article-pdf/417/1/472/3026193/mnras0417-0472.pdf},
}

@article{10.1111/j.1365-2966.2010.17409.x,
    author = {Donati, J.-F. and Skelly, M. B. and Bouvier, J. and Gregory, S. G. and Grankin, K. N. and Jardine, M. M. and Hussain, G. A. J. and M{\'{e}}nard, F. and Dougados, C. and Unruh, Y. and Mohanty, S. and Auri{\`{e}}re, M. and Morin, J. and Far{\`{e}}s, R.},
    title = {Magnetospheric accretion and spin-down of the prototypical classical T Tauri star AA Tau},
    journal = {Monthly Notices of the Royal Astronomical Society},
    volume = {409},
    number = {4},
    pages = {1347-1361},
    year = {2010},
    month = {12},
    issn = {0035-8711},
    doi = {10.1111/j.1365-2966.2010.17409.x},
    url = {https://doi.org/10.1111/j.1365-2966.2010.17409.x},
    eprint = {https://academic.oup.com/mnras/article-pdf/409/4/1347/3070562/mnras0409-1347.pdf},
}

@article{10.1111/j.1365-2966.2008.13111.x,
    author = {Donati, J.-F. and Jardine, M. M. and Gregory, S. G. and Petit, P. and Paletou, F. and Bouvier, J. and Dougados, C. and M{\'{e}}nard, F. and Cameron, A. C. and Harries, T. J. and Hussain, G. A. J. and Unruh, Y. and Morin, J. and Marsden, S. C. and Manset, N. and Auri{\`{e}}re, M. and Catala, C. and Alecian, E.},
    title = {Magnetospheric accretion on the T Tauri star BP Tauri},
    journal = {Monthly Notices of the Royal Astronomical Society},
    volume = {386},
    number = {3},
    pages = {1234-1251},
    year = {2008},
    month = {05},
    issn = {0035-8711},
    doi = {10.1111/j.1365-2966.2008.13111.x},
    url = {https://doi.org/10.1111/j.1365-2966.2008.13111.x},
    eprint = {https://academic.oup.com/mnras/article-pdf/386/3/1234/3615549/mnras0386-1234.pdf},
}

@article{10.1111/j.1365-2966.2007.12194.x,
    author = {Donati, J.-F. and Jardine, M. M. and Gregory, S. G. and Petit, P. and Bouvier, J. and Dougados, C. and M{\'{e}}nard, F. and Cameron, A. C. and Harries, T. J. and Jeffers, S.V. and Paletou, F.},
    title = {Magnetic fields and accretion flows on the classical T Tauri star V2129 Oph},
    journal = {Monthly Notices of the Royal Astronomical Society},
    volume = {380},
    number = {4},
    pages = {1297-1312},
    year = {2007},
    month = {09},
    issn = {0035-8711},
    doi = {10.1111/j.1365-2966.2007.12194.x},
    url = {https://doi.org/10.1111/j.1365-2966.2007.12194.x},
    eprint = {https://academic.oup.com/mnras/article-pdf/380/4/1297/18688451/mnras0380-1297.pdf},
}

@ARTICLE{2025ApJ...989....5T,
       author = {{Trapman}, Leon and {Zhang}, Ke and {Rosotti}, Giovanni P. and {Pinilla}, Paola and {Tabone}, Beno{\^\i}t and {Pascucci}, Ilaria and {Agurto-Gangas}, Carolina and {Anania}, Rossella and {Carpenter}, John and {Cieza}, Lucas A. and {Deng}, Dingshan and {Gonz{\'a}lez-Ruilova}, Camilo and {Hogerheijde}, Michiel R. and {Kurtovic}, Nicol{\'a}s T. and {Kuznetsova}, Aleksandra and {Miley}, James and {P{\'e}rez}, Laura M. and {Ruiz-Rodriguez}, Dary A. and {Schwarz}, Kamber and {Sierra}, Anibal and {TorresVillanueva}, Estephani and {Vioque}, Miguel},
        title = "{The ALMA Survey of Gas Evolution of PROtoplanetary Disks (AGE-PRO). V. Protoplanetary Gas Disk Masses}",
      journal = {\apj},
         year = 2025,
        month = aug,
       volume = {989},
       number = {1},
          eid = {5},
        pages = {5},
          doi = {10.3847/1538-4357/adcd6e},
archivePrefix = {arXiv},
       eprint = {2506.10738},
 primaryClass = {astro-ph.EP},
       adsurl = {https://ui.adsabs.harvard.edu/abs/2025ApJ...989....5T}
}

@ARTICLE{2009ARA&A..47..481A,
       author = {{Asplund}, Martin and {Grevesse}, Nicolas and {Sauval}, A. Jacques and {Scott}, Pat},
        title = "{The Chemical Composition of the Sun}",
      journal = {\araa},
         year = 2009,
        month = sep,
       volume = {47},
       number = {1},
        pages = {481-522},
          doi = {10.1146/annurev.astro.46.060407.145222},
archivePrefix = {arXiv},
       eprint = {0909.0948},
 primaryClass = {astro-ph.SR},
       adsurl = {https://ui.adsabs.harvard.edu/abs/2009ARA&A..47..481A}
}

@ARTICLE{2008A&A...480..889S,
       author = {{Santos}, N.~C. and {Melo}, C. and {James}, D.~J. and {Gameiro}, J.~F. and {Bouvier}, J. and {Gomes}, J.~I.},
        title = "{Chemical abundances in six nearby star-forming regions. Implications for galactic evolution and planet searches around very young stars}",
      journal = {\aap},
         year = 2008,
        month = mar,
       volume = {480},
       number = {3},
        pages = {889-897},
          doi = {10.1051/0004-6361:20079083},
archivePrefix = {arXiv},
       eprint = {0801.2529},
 primaryClass = {astro-ph},
       adsurl = {https://ui.adsabs.harvard.edu/abs/2008A&A...480..889S}
}

@ARTICLE{2011A&A...526A.103D,
       author = {{D'Orazi}, V. and {Biazzo}, K. and {Randich}, S.},
        title = "{Chemical composition of the Taurus-Auriga association}",
      journal = {\aap},
         year = 2011,
        month = feb,
       volume = {526},
          eid = {A103},
        pages = {A103},
          doi = {10.1051/0004-6361/201015616},
archivePrefix = {arXiv},
       eprint = {1012.0848},
 primaryClass = {astro-ph.SR},
       adsurl = {https://ui.adsabs.harvard.edu/abs/2011A&A...526A.103D}
}

@INPROCEEDINGS{2023ASPC..534..539M,
       author = {{Manara}, C.~F. and {Ansdell}, M. and {Rosotti}, G.~P. and {Hughes}, A.~M. and {Armitage}, P.~J. and {Lodato}, G. and {Williams}, J.~P.},
        title = "{Demographics of Young Stars and their Protoplanetary Disks: Lessons Learned on Disk Evolution and its Connection to Planet Formation}",
    booktitle = {Protostars and Planets VII},
         year = 2023,
       editor = {{Inutsuka}, S. and {Aikawa}, Y. and {Muto}, T. and {Tomida}, K. and {Tamura}, M.},
       series = {Astronomical Society of the Pacific Conference Series},
       volume = {534},
        month = jul,
        pages = {539},
          doi = {10.48550/arXiv.2203.09930},
archivePrefix = {arXiv},
       eprint = {2203.09930},
 primaryClass = {astro-ph.SR},
       adsurl = {https://ui.adsabs.harvard.edu/abs/2023ASPC..534..539M}
}

@ARTICLE{2025A&A...693A.302S,
       author = {{Savvidou}, Sofia and {Bitsch}, Bertram},
        title = "{There is no disk mass budget problem of planet formation}",
      journal = {\aap},
         year = 2025,
        month = jan,
       volume = {693},
          eid = {A302},
        pages = {A302},
          doi = {10.1051/0004-6361/202449847},
archivePrefix = {arXiv},
       eprint = {2407.08533},
 primaryClass = {astro-ph.EP},
       adsurl = {https://ui.adsabs.harvard.edu/abs/2025A&A...693A.302S}
}

@ARTICLE{2009ApJ...690.1193S,
       author = {{Sargent}, B.~A. and {Forrest}, W.~J. and {Tayrien}, C. and {McClure}, M.~K. and {Li}, A. and {Basu}, A.~R. and {Manoj}, P. and {Watson}, D.~M. and {Bohac}, C.~J. and {Furlan}, E. and {Kim}, K.~H. and {Green}, J.~D. and {Sloan}, G.~C.},
        title = "{Silica in Protoplanetary Disks}",
      journal = {\apj},
         year = 2009,
        month = jan,
       volume = {690},
       number = {2},
        pages = {1193-1207},
          doi = {10.1088/0004-637X/690/2/1193},
archivePrefix = {arXiv},
       eprint = {0811.3590},
 primaryClass = {astro-ph},
       adsurl = {https://ui.adsabs.harvard.edu/abs/2009ApJ...690.1193S}
}

@ARTICLE{2015ApJ...801L...7R,
       author = {{Roskosz}, Mathieu and {Leroux}, Hugues},
        title = "{A Significant Amount of Crystalline Silica in Returned Cometary Samples: Bridging the Gap between Astrophysical and Meteoritical Observations}",
      journal = {\apjl},
         year = 2015,
        month = mar,
       volume = {801},
       number = {1},
          eid = {L7},
        pages = {L7},
          doi = {10.1088/2041-8205/801/1/L7},
       adsurl = {https://ui.adsabs.harvard.edu/abs/2015ApJ...801L...7R}
}

@ARTICLE{2023A&A...679A..42S,
       author = {{Savvidou}, Sofia and {Bitsch}, Bertram},
        title = "{How to make giant planets via pebble accretion}",
      journal = {\aap},
         year = 2023,
        month = nov,
       volume = {679},
          eid = {A42},
        pages = {A42},
          doi = {10.1051/0004-6361/202245793},
archivePrefix = {arXiv},
       eprint = {2309.03807},
 primaryClass = {astro-ph.EP},
       adsurl = {https://ui.adsabs.harvard.edu/abs/2023A&A...679A..42S}
}

@ARTICLE{2018A&A...618A.132K,
       author = {{Kunitomo}, Masanobu and {Guillot}, Tristan and {Ida}, Shigeru and {Takeuchi}, Taku},
        title = "{Revisiting the pre-main-sequence evolution of stars. II. Consequences of planet formation on stellar surface composition}",
      journal = {\aap},
         year = 2018,
        month = oct,
       volume = {618},
          eid = {A132},
        pages = {A132},
          doi = {10.1051/0004-6361/201833127},
archivePrefix = {arXiv},
       eprint = {1808.07396},
 primaryClass = {astro-ph.SR},
       adsurl = {https://ui.adsabs.harvard.edu/abs/2018A&A...618A.132K}
}

@ARTICLE{2018MNRAS.476.4418J,
       author = {{Jermyn}, Adam S. and {Kama}, Mihkel},
        title = "{Stellar photospheric abundances as a probe of discs and planets}",
      journal = {\mnras},
         year = 2018,
        month = jun,
       volume = {476},
       number = {4},
        pages = {4418-4434},
          doi = {10.1093/mnras/sty429},
archivePrefix = {arXiv},
       eprint = {1804.06414},
 primaryClass = {astro-ph.SR},
       adsurl = {https://ui.adsabs.harvard.edu/abs/2018MNRAS.476.4418J}
}

@ARTICLE{2019ApJ...882...49L,
       author = {{Long}, Feng and {Herczeg}, Gregory J. and {Harsono}, Daniel and {Pinilla}, Paola and {Tazzari}, Marco and {Manara}, Carlo F. and {Pascucci}, Ilaria and {Cabrit}, Sylvie and {Nisini}, Brunella and {Johnstone}, Doug and {Edwards}, Suzan and {Salyk}, Colette and {Menard}, Francois and {Lodato}, Giuseppe and {Boehler}, Yann and {Mace}, Gregory N. and {Liu}, Yao and {Mulders}, Gijs D. and {Hendler}, Nathanial and {Ragusa}, Enrico and {Fischer}, William J. and {Banzatti}, Andrea and {Rigliaco}, Elisabetta and {van de Plas}, Gerrit and {Dipierro}, Giovanni and {Gully-Santiago}, Michael and {Lopez-Valdivia}, Ricardo},
        title = "{Compact Disks in a High-resolution ALMA Survey of Dust Structures in the Taurus Molecular Cloud}",
      journal = {\apj},
         year = 2019,
        month = sep,
       volume = {882},
       number = {1},
          eid = {49},
        pages = {49},
          doi = {10.3847/1538-4357/ab2d2d},
archivePrefix = {arXiv},
       eprint = {1906.10809},
 primaryClass = {astro-ph.SR},
       adsurl = {https://ui.adsabs.harvard.edu/abs/2019ApJ...882...49L}
}

@ARTICLE{1994ApJ...426..669H,
       author = {{Hartmann}, Lee and {Hewett}, Robert and {Calvet}, Nuria},
        title = "{Magnetospheric Accretion Models for T Tauri Stars. I. Balmer Line Profiles without Rotation}",
      journal = {\apj},
         year = 1994,
        month = may,
       volume = {426},
        pages = {669},
          doi = {10.1086/174104},
       adsurl = {https://ui.adsabs.harvard.edu/abs/1994ApJ...426..669H}
}

@ARTICLE{1976ApJS...30....1V,
       author = {{Vernazza}, J.~E. and {Avrett}, E.~H. and {Loeser}, R.},
        title = "{Structure of the solar chromosphere. II. The underlying photosphere and temperature-minimum region.}",
      journal = {\apjs},
         year = 1976,
        month = jan,
       volume = {30},
        pages = {1-60},
          doi = {10.1086/190356},
       adsurl = {https://ui.adsabs.harvard.edu/abs/1976ApJS...30....1V}
}

@article{Patino_2026,
doi = {10.3847/1538-4357/ae5d2e},
url = {https://doi.org/10.3847/1538-4357/ae5d2e},
year = {2026},
month = {may},
publisher = {The American Astronomical Society},
volume = {1002},
number = {2},
pages = {190},
author = {Pati{\~{n}}o, Naiara and Calvet, Nuria and Magris, Gladis and Micolta, Marbely and Thanathibodee, Thanawuth and Waters, Thomas K. and Colmenares, Mar{\'{i}}a Jos{\'{e}}},
title = {Understanding Balmer Decrements in T Tauri Stars in Terms of Multiflow Magnetospheric Accretion},
journal = {The Astrophysical Journal}
}

@article{Thanathibodee_2023,
doi = {10.3847/1538-4357/acac84},
url = {https://doi.org/10.3847/1538-4357/acac84},
year = {2023},
month = {feb},
publisher = {The American Astronomical Society},
volume = {944},
number = {1},
pages = {90},
author = {Thanathibodee, Thanawuth and Molina, Brandon and Serna, Javier and Calvet, Nuria and Hern{\'{a}}ndez, Jes{\'{u}}s and Muzerolle, James and Franco-Hern{\'{a}}ndez, Ramiro},
title = {A Census of the Low Accretors. II. Accretion Properties},
journal = {The Astrophysical Journal}
}

@article{Muzerolle_2001,
doi = {10.1086/319779},
url = {https://doi.org/10.1086/319779},
year = {2001},
month = {apr},
publisher = {The American Astronomical Society},
volume = {550},
number = {2},
pages = {944},
author = {Muzerolle, James and Calvet,, Nuria and Hartmann, Lee},
title = {Emission-Line Diagnostics of T Tauri Magnetospheric
Accretion. II. Improved
Model Tests and Insights into Accretion Physics},
journal = {The Astrophysical Journal}
}

@ARTICLE{2025A&A...694A.147G,
       author = {{Gasman}, Danny and {Temmink}, Milou and {van Dishoeck}, Ewine F. and {Kurtovic}, Nicolas T. and {Grant}, Sierra L. and {Sellek}, Andrew and {Tabone}, Beno{\^\i}t and {Henning}, Thomas and {Kamp}, Inga and {G{\"u}del}, Manuel and {Barrado}, David and {Caratti o Garatti}, Alessio and {Glauser}, Adrian M. and {Waters}, Laurens B.~F.~M. and {Arabhavi}, Aditya M. and {Jang}, Hyerin and {Kanwar}, Jayatee and {Lienert}, Julia L. and {Perotti}, Giulia and {Schwarz}, Kamber and {Vlasblom}, Marissa},
        title = "{MINDS: The influence of outer dust disc structure on the volatile delivery to the inner disc}",
      journal = {\aap},
         year = 2025,
        month = feb,
       volume = {694},
          eid = {A147},
        pages = {A147},
          doi = {10.1051/0004-6361/202452152},
archivePrefix = {arXiv},
       eprint = {2501.04587},
 primaryClass = {astro-ph.EP},
       adsurl = {https://ui.adsabs.harvard.edu/abs/2025A&A...694A.147G}
}

@ARTICLE{2023A&A...671A.140G,
       author = {{Guzm{\'a}n-D{\'\i}az}, J. and {Montesinos}, B. and {Mendigut{\'\i}a}, I. and {Kama}, M. and {Meeus}, G. and {Vioque}, M. and {Oudmaijer}, R.~D. and {Villaver}, E.},
        title = "{Relation between metallicities and spectral energy distributions of Herbig Ae/Be stars. A potential link with planet formation}",
      journal = {\aap},
         year = 2023,
        month = mar,
       volume = {671},
          eid = {A140},
        pages = {A140},
          doi = {10.1051/0004-6361/202245427},
archivePrefix = {arXiv},
       eprint = {2212.14022},
 primaryClass = {astro-ph.SR},
       adsurl = {https://ui.adsabs.harvard.edu/abs/2023A&A...671A.140G}
}

@ARTICLE{2005ApJ...630L.185C,
       author = {{Calvet}, N. and {D'Alessio}, P. and {Watson}, D.~M. and {Franco-Hern{\'a}ndez}, R. and {Furlan}, E. and {Green}, J. and {Sutter}, P.~M. and {Forrest}, W.~J. and {Hartmann}, L. and {Uchida}, K.~I. and {Keller}, L.~D. and {Sargent}, B. and {Najita}, J. and {Herter}, T.~L. and {Barry}, D.~J. and {Hall}, P.},
        title = "{Disks in Transition in the Taurus Population: Spitzer IRS Spectra of GM Aurigae and DM Tauri}",
      journal = {\apjl},
         year = 2005,
        month = sep,
       volume = {630},
       number = {2},
        pages = {L185-L188},
          doi = {10.1086/491652},
       adsurl = {https://ui.adsabs.harvard.edu/abs/2005ApJ...630L.185C}
}

@article{Ingleby_2011,
doi = {10.1088/0004-637X/743/2/105},
url = {https://doi.org/10.1088/0004-637X/743/2/105},
year = {2011},
month = {nov},
publisher = {The American Astronomical Society},
volume = {743},
number = {2},
pages = {105},
author = {Ingleby, Laura and Calvet, Nuria and Bergin, Edwin and Herczeg, Gregory and Brown, Alexander and Alexander, Richard and Edwards, Suzan and Espaillat, Catherine and France, Kevin and Gregory, Scott G. and Hillenbrand, Lynne and Roueff, Evelyne and Valenti, Jeff and Walter, Frederick and Johns-Krull, Christopher and Brown, Joanna and Linsky, Jeffrey and McClure, Melissa and Ardila, David and Abgrall, Herv{\'{e}} and Bethell, Thomas and Hussain, Gaitee and Yang, Hao},
title = {NEAR-ULTRAVIOLET EXCESS IN SLOWLY ACCRETING T TAURI STARS: LIMITS IMPOSED BY CHROMOSPHERIC EMISSION*},
journal = {The Astrophysical Journal}
}

@ARTICLE{2007ApJ...670L.135E,
       author = {{Espaillat}, C. and {Calvet}, N. and {D'Alessio}, P. and {Hern{\'a}ndez}, J. and {Qi}, C. and {Hartmann}, L. and {Furlan}, E. and {Watson}, D.~M.},
       title = "{On the Diversity of the Taurus Transitional Disks: UX Tauri A and LkCa 15}",
       journal = {\apjl},
       year = 2007,
       month = dec,
       volume = {670},
       number = {2},
       pages = {L135-L138},
       doi = {10.1086/524360},
       archivePrefix = {arXiv},
       eprint = {0710.2892},
       primaryClass = {astro-ph},
       adsurl = {https://ui.adsabs.harvard.edu/abs/2007ApJ...670L.135E}
       }

@ARTICLE{2020SciPy-NMeth,
  author  = {Virtanen, Pauli and Gommers, Ralf and Oliphant, Travis E. and
            Haberland, Matt and Reddy, Tyler and Cournapeau, David and
            Burovski, Evgeni and Peterson, Pearu and Weckesser, Warren and
            Bright, Jonathan and {van der Walt}, St{\'e}fan J. and
            Brett, Matthew and Wilson, Joshua and Millman, K. Jarrod and
            Mayorov, Nikolay and Nelson, Andrew R. J. and Jones, Eric and
            Kern, Robert and Larson, Eric and Carey, C J and
            Polat, {\.I}lhan and Feng, Yu and Moore, Eric W. and
            {VanderPlas}, Jake and Laxalde, Denis and Perktold, Josef and
            Cimrman, Robert and Henriksen, Ian and Quintero, E. A. and
            Harris, Charles R. and Archibald, Anne M. and
            Ribeiro, Ant{\^o}nio H. and Pedregosa, Fabian and
            {van Mulbregt}, Paul and {SciPy 1.0 Contributors}},
  title   = {{{SciPy} 1.0: Fundamental Algorithms for Scientific
            Computing in Python}},
  journal = {Nature Methods},
  year    = {2020},
  volume  = {17},
  pages   = {261--272},
  adsurl  = {https://rdcu.be/b08Wh},
  doi     = {10.1038/s41592-019-0686-2},
}

@article{Spina_2014,
  title         = {The<i>Gaia</i>-ESO Survey: Metallicity of the Chamaeleon I star-forming region},
  author        = {Spina, L. and Randich, S. and Palla, F. and Biazzo, K. and Sacco, G. G. and Alfaro, E. J. and Franciosini, E. and Magrini, L. and Morbidelli, L. and Frasca, A. and Adibekyan, V. and Delgado-Mena, E. and Sousa, S. G. and Gonz{\'{a}}lez Hern{\'{a}}ndez, J. I. and Montes, D. and Tabernero, H. and Tautvai{\v{s}}ien{\.{e}}, G. and Bonito, R. and Lanzafame, A. C. and Gilmore, G. and Jeffries, R. D. and Vallenari, A. and Bensby, T. and Bragaglia, A. and Flaccomio, E. and Korn, A. J. and Pancino, E. and Recio-Blanco, A. and Smiljanic, R. and Bergemann, M. and Costado, M. T. and Damiani, F. and Hill, V. and Hourihane, A. and Jofr{\'{e}}, P. and de Laverny, P. and Lardo, C. and Masseron, T. and Prisinzano, L. and Worley, C. C.},
  year          = {2014},
  month         = Aug,
  journal       = {Astronomy \& Astrophysics},
  publisher     = {EDP Sciences},
  volume        = {568},
  pages         = {A2},
  doi           = {10.1051/0004-6361/201424135},
  issn          = {1432-0746},
  url           = {http://dx.doi.org/10.1051/0004-6361/201424135}
}

@inproceedings{clemens_goodman_2004,
	booktitle = {Ground-based Instrumentation for Astronomy},
	series = {Proceedings of SPIE},
	volume = {5492},
	address = {USA},
	title = {The {Goodman} spectrograph},
	url = {http://proceedings.spiedigitallibrary.org/proceeding.aspx?doi=10.1117/12.550069},
	doi = {10.1117/12.550069},
	urldate = {2026-04-23},
	author = {Clemens, J. Christopher and Crain, J. Adam and Anderson, Robert},
	month = sep,
	year = {2004},
	pages = {331},
}

@inproceedings{Bernstein_2003,
  title         = {MIKE: A Double Echelle Spectrograph for the Magellan Telescopes at Las Campanas Observatory},
  author        = {Bernstein, Rebecca and Shectman, Stephen A. and Gunnels, Steven M. and Mochnacki, Stefan and Athey, Alex E.},
  year          = {2003},
  month         = Mar,
  booktitle     = {Instrument Design and Performance for Optical/Infrared Ground-based Telescopes},
  publisher     = {SPIE},
  volume        = {4841},
  pages         = {1694},
  doi           = {10.1117/12.461502},
  issn          = {0277-786X},
  url           = {http://dx.doi.org/10.1117/12.461502},
  editor        = {Iye, Masanori and Moorwood, Alan F. M.}
}

@article{Micolta_2024,
  title         = {Using the Ca ii Lines in T Tauri Stars to Infer the Abundance of Refractory Elements in the Innermost Disk Region},
  author        = {Micolta, Marbely and Calvet, Nuria and Thanathibodee, Thanawuth and Magris C., Gladis and Manara, Carlo F. and Venuti, Laura and Alcal{\'{a}}, Juan Manuel and Herczeg, Gregory J.},
  year          = {2024},
  month         = nov,
  journal       = {The Astrophysical Journal},
  publisher     = {American Astronomical Society},
  volume        = {976},
  number        = {2},
  pages         = {251},
  doi           = {10.3847/1538-4357/ad8884},
  issn          = {1538-4357},
  url           = {http://dx.doi.org/10.3847/1538-4357/ad8884}
}

@article{Yamaguchi_2024,
  title         = {ALMA 2D super-resolution imaging of Taurus--Auriga protoplanetary disks: Probing statistical properties of disk substructures},
  author        = {Yamaguchi, Masayuki and Muto, Takayuki and Tsukagoshi, Takashi and Nomura, Hideko and Hirano, Naomi and Nakazato, Takeshi and Ikeda, Shiro and Tamura, Motohide and Kawabe, Ryohei},
  year          = {2024},
  month         = may,
  journal       = {Publications of the Astronomical Society of Japan},
  publisher     = {Oxford University Press (OUP)},
  volume        = {76},
  number        = {3},
  pages         = {437--474},
  doi           = {10.1093/pasj/psae022},
  issn          = {2053-051X},
  url           = {http://dx.doi.org/10.1093/pasj/psae022}
}

@article{Zhang_2023,
  title         = {Substructures in Compact Disks of the Taurus Star-forming Region},
  author        = {Zhang, Shangjia and Kalscheur, Matt and Long, Feng and Zhang, Ke and Long, Deryl E. and Bergin, Edwin A. and Zhu, Zhaohuan and Trapman, Leon},
  year          = {2023},
  month         = jul,
  journal       = {The Astrophysical Journal},
  publisher     = {American Astronomical Society},
  volume        = {952},
  number        = {2},
  pages         = {108},
  doi           = {10.3847/1538-4357/acd334},
  issn          = {1538-4357},
  url           = {http://dx.doi.org/10.3847/1538-4357/acd334}
}

@article{McClure_2020,
  title         = {Measuring the atomic composition of planetary building blocks},
  author        = {McClure, M. K. and Dominik, C. and Kama, M.},
  year          = {2020},
  month         = oct,
  journal       = {Astronomy \& Astrophysics},
  publisher     = {EDP Sciences},
  volume        = {642},
  pages         = {L15},
  doi           = {10.1051/0004-6361/202038912},
  issn          = {1432-0746},
  url           = {http://dx.doi.org/10.1051/0004-6361/202038912}
}

@article{McClure_2019,
  title         = {Carbon depletion observed inside T Tauri inner rims: Formation of icy, kilometer size planetesimals by 1 Myr},
  author        = {McClure, M. K.},
  year          = {2019},
  month         = nov,
  journal       = {Astronomy \& Astrophysics},
  publisher     = {EDP Sciences},
  volume        = {632},
  pages         = {A32},
  doi           = {10.1051/0004-6361/201834361},
  issn          = {1432-0746},
  url           = {http://dx.doi.org/10.1051/0004-6361/201834361}
}

@article{Kama_2015,
  title         = {Fingerprints of giant planets in the photospheres of Herbig stars},
  author        = {Kama, M. and Folsom, C. P. and Pinilla, P.},
  year          = {2015},
  month         = oct,
  journal       = {Astronomy \& Astrophysics},
  publisher     = {EDP Sciences},
  volume        = {582},
  pages         = {L10},
  doi           = {10.1051/0004-6361/201527094},
  issn          = {1432-0746},
  url           = {http://dx.doi.org/10.1051/0004-6361/201527094}
}

@ARTICLE{2009ApJS..180...84W,
       author = {{Watson}, Dan M. and {Leisenring}, Jarron M. and {Furlan}, Elise and {Bohac}, C.~J. and {Sargent}, B. and {Forrest}, W.~J. and {Calvet}, Nuria and {Hartmann}, Lee and {Nordhaus}, Jason T. and {Green}, Joel D. and {Kim}, K.~H. and {Sloan}, G.~C. and {Chen}, C.~H. and {Keller}, L.~D. and {d'Alessio}, Paola and {Najita}, J. and {Uchida}, Keven I. and {Houck}, J.~R.},
        title = "{Crystalline Silicates and Dust Processing in the Protoplanetary Disks of the Taurus Young Cluster}",
      journal = {\apjs},
         year = 2009,
        month = jan,
       volume = {180},
       number = {1},
        pages = {84-101},
          doi = {10.1088/0067-0049/180/1/84},
archivePrefix = {arXiv},
       eprint = {0704.1518},
 primaryClass = {astro-ph},
       adsurl = {https://ui.adsabs.harvard.edu/abs/2009ApJS..180...84W}
}

@INPROCEEDINGS{2014prpl.conf..497E,
       author = {{Espaillat}, C. and {Muzerolle}, J. and {Najita}, J. and {Andrews}, S. and {Zhu}, Z. and {Calvet}, N. and {Kraus}, S. and {Hashimoto}, J. and {Kraus}, A. and {D'Alessio}, P.},
        title = "{An Observational Perspective of Transitional Disks}",
    booktitle = {Protostars and Planets VI},
         year = 2014,
       editor = {{Beuther}, Henrik and {Klessen}, Ralf S. and {Dullemond}, Cornelis P. and {Henning}, Thomas},
        month = jan,
        pages = {497-520},
          doi = {10.2458/azu_uapress_9780816531240-ch022},
archivePrefix = {arXiv},
       eprint = {1402.7103},
 primaryClass = {astro-ph.SR},
       adsurl = {https://ui.adsabs.harvard.edu/abs/2014prpl.conf..497E}
}

@ARTICLE{2011A&A...536A.105V,
       author = {{Vernet}, J. and {Dekker}, H. and {D'Odorico}, S. and {Kaper}, L. and {Kjaergaard}, P. and {Hammer}, F. and {Randich}, S. and {Zerbi}, F. and {Groot}, P.~J. and {Hjorth}, J. and {Guinouard}, I. and {Navarro}, R. and {Adolfse}, T. and {Albers}, P.~W. and {Amans}, J.-P. and {Andersen}, J.~J. and {Andersen}, M.~I. and {Binetruy}, P. and {Bristow}, P. and {Castillo}, R. and {Chemla}, F. and {Christensen}, L. and {Conconi}, P. and {Conzelmann}, R. and {Dam}, J. and {de Caprio}, V. and {de Ugarte Postigo}, A. and {Delabre}, B. and {di Marcantonio}, P. and {Downing}, M. and {Elswijk}, E. and {Finger}, G. and {Fischer}, G. and {Flores}, H. and {Fran{\c{c}}ois}, P. and {Goldoni}, P. and {Guglielmi}, L. and {Haigron}, R. and {Hanenburg}, H. and {Hendriks}, I. and {Horrobin}, M. and {Horville}, D. and {Jessen}, N.~C. and {Kerber}, F. and {Kern}, L. and {Kiekebusch}, M. and {Kleszcz}, P. and {Klougart}, J. and {Kragt}, J. and {Larsen}, H.~H. and {Lizon}, J.-L. and {Lucuix}, C. and {Mainieri}, V. and {Manuputy}, R. and {Martayan}, C. and {Mason}, E. and {Mazzoleni}, R. and {Michaelsen}, N. and {Modigliani}, A. and {Moehler}, S. and {M{\o}ller}, P. and {Norup S{\o}rensen}, A. and {N{\o}rregaard}, P. and {P{\'e}roux}, C. and {Patat}, F. and {Pena}, E. and {Pragt}, J. and {Reinero}, C. and {Rigal}, F. and {Riva}, M. and {Roelfsema}, R. and {Royer}, F. and {Sacco}, G. and {Santin}, P. and {Schoenmaker}, T. and {Spano}, P. and {Sweers}, E. and {Ter Horst}, R. and {Tintori}, M. and {Tromp}, N. and {van Dael}, P. and {van der Vliet}, H. and {Venema}, L. and {Vidali}, M. and {Vinther}, J. and {Vola}, P. and {Winters}, R. and {Wistisen}, D. and {Wulterkens}, G. and {Zacchei}, A.},
        title = "{X-shooter, the new wide band intermediate resolution spectrograph at the ESO Very Large Telescope}",
      journal = {\aap},
         year = 2011,
        month = dec,
       volume = {536},
          eid = {A105},
        pages = {A105},
          doi = {10.1051/0004-6361/201117752},
archivePrefix = {arXiv},
       eprint = {1110.1944},
 primaryClass = {astro-ph.IM},
       adsurl = {https://ui.adsabs.harvard.edu/abs/2011A&A...536A.105V}
}

@ARTICLE{2013ApJ...767..112I,
       author = {{Ingleby}, Laura and {Calvet}, Nuria and {Herczeg}, Gregory and {Blaty}, Alex and {Walter}, Frederick and {Ardila}, David and {Alexander}, Richard and {Edwards}, Suzan and {Espaillat}, Catherine and {Gregory}, Scott G. and {Hillenbrand}, Lynne and {Brown}, Alexander},
        title = "{Accretion Rates for T Tauri Stars Using Nearly Simultaneous Ultraviolet and Optical Spectra}",
      journal = {\apj},
         year = 2013,
        month = apr,
       volume = {767},
       number = {2},
          eid = {112},
        pages = {112},
          doi = {10.1088/0004-637X/767/2/112},
archivePrefix = {arXiv},
       eprint = {1303.0769},
 primaryClass = {astro-ph.SR},
       adsurl = {https://ui.adsabs.harvard.edu/abs/2013ApJ...767..112I}
}

@ARTICLE{2001ApJ...553..321D,
       author = {{D'Alessio}, Paola and {Calvet}, Nuria and {Hartmann}, Lee},
        title = "{Accretion Disks around Young Objects. III. Grain Growth}",
      journal = {\apj},
         year = 2001,
        month = may,
       volume = {553},
       number = {1},
        pages = {321-334},
          doi = {10.1086/320655},
archivePrefix = {arXiv},
       eprint = {astro-ph/0101443},
 primaryClass = {astro-ph},
       adsurl = {https://ui.adsabs.harvard.edu/abs/2001ApJ...553..321D}
}

@ARTICLE{2009ApJS..182..477S,
       author = {{Sargent}, B.~A. and {Forrest}, W.~J. and {Tayrien}, C. and {McClure}, M.~K. and {Watson}, Dan M. and {Sloan}, G.~C. and {Li}, A. and {Manoj}, P. and {Bohac}, C.~J. and {Furlan}, E. and {Kim}, K.~H. and {Green}, J.~D.},
        title = "{Dust Processing and Grain Growth in Protoplanetary Disks in the Taurus-Auriga Star-Forming Region}",
      journal = {\apjs},
         year = 2009,
        month = jun,
       volume = {182},
       number = {2},
        pages = {477-508},
          doi = {10.1088/0067-0049/182/2/477},
archivePrefix = {arXiv},
       eprint = {0811.3622},
 primaryClass = {astro-ph},
       adsurl = {https://ui.adsabs.harvard.edu/abs/2009ApJS..182..477S}
}

@misc{2021ascl.soft04010D,
       author = {{Dominik}, Carsten and {Min}, Michiel and {Tazaki}, Ryo},
        title = "{OpTool: Command-line driven tool for creating complex dust opacities}",
 howpublished = {Astrophysics Source Code Library, record ascl:2104.010},
         year = 2021,
        month = apr,
          eid = {ascl:2104.010},
archivePrefix = {ascl},
       eprint = {2104.010},
       adsurl = {https://ui.adsabs.harvard.edu/abs/2021ascl.soft04010D}
}

@ARTICLE{2000ApJ...531..159K,
       author = {{Kelson}, Daniel D. and {Illingworth}, Garth D. and {van Dokkum}, Pieter G. and {Franx}, Marijn},
        title = "{The Evolution of Early-Type Galaxies in Distant Clusters. II. Internal Kinematics of 55 Galaxies in the z=0.33 Cluster CL 1358+62}",
      journal = {\apj},
         year = 2000,
        month = mar,
       volume = {531},
       number = {1},
        pages = {159-183},
          doi = {10.1086/308445},
archivePrefix = {arXiv},
       eprint = {astro-ph/9908257},
 primaryClass = {astro-ph},
       adsurl = {https://ui.adsabs.harvard.edu/abs/2000ApJ...531..159K}
}

@article{van_der_Marel_2021,
    doi = {10.3847/1538-3881/ac0255},
    url = {http://dx.doi.org/10.3847/1538-3881/ac0255},
    issn = {1538-3881},
    year = {2021},
    month = {jun},
    pages = {28},
    title = {A Stellar Mass Dependence of Structured Disks: A Possible Link with Exoplanet Demographics},
    author = {van der Marel, Nienke and Mulders, Gijs D.},
    number = {1},
    volume = {162},
    journal = {The Astronomical Journal},
    publisher = {American Astronomical Society}}

@article{Zhu_2012,
    doi = {10.1088/0004-637x/746/1/110},
    url = {http://dx.doi.org/10.1088/0004-637X/746/1/110},
    issn = {1538-4357},
    year = {2012},
    month = {jan},
    pages = {110},
    title = {CHALLENGES IN FORMING PLANETS BY GRAVITATIONAL INSTABILITY: DISK IRRADIATION AND CLUMP MIGRATION, ACCRETION, AND TIDAL DESTRUCTION},
    author = {Zhu, Zhaohuan and Hartmann, Lee and Nelson, Richard P. and Gammie, Charles F.},
    number = {1},
    volume = {746},
    journal = {The Astrophysical Journal},
    publisher = {American Astronomical Society}}

@article{Pinilla_2012,
    doi = {10.1051/0004-6361/201118204},
    url = {http://dx.doi.org/10.1051/0004-6361/201118204},
    issn = {1432-0746},
    year = {2012},
    month = {feb},
    pages = {A114},
    title = {Trapping dust particles in the outer regions of protoplanetary disks},
    author = {Pinilla, P. and Birnstiel, T. and Ricci, L. and Dullemond, C. P. and Uribe, A. L. and Testi, L. and Natta, A.},
    volume = {538},
    journal = {Astronomy \& Astrophysics},
    publisher = {EDP Sciences}}

@article{lambrechts_rapid_2012,
	title = {Rapid growth of gas-giant cores by pebble accretion},
	volume = {544},
	issn = {1432-0746},
	url = {http://dx.doi.org/10.1051/0004-6361/201219127},
	doi = {10.1051/0004-6361/201219127},
	journal = {Astronomy \& Astrophysics},
	author = {Lambrechts, M. and Johansen, A.},
	month = jul,
	year = {2012},
	pages = {A32},
}

@article{birnstiel_simple_2012,
	title = {A simple model for the evolution of the dust population in protoplanetary disks},
	volume = {539},
	issn = {1432-0746},
	url = {http://dx.doi.org/10.1051/0004-6361/201118136},
	doi = {10.1051/0004-6361/201118136},
	journal = {Astronomy \& Astrophysics},
	author = {Birnstiel, T. and Klahr, H. and Ercolano, B.},
	month = mar,
	year = {2012},
	pages = {A148},
}

@article{huhn_how_2023,
	title = {How accretion of planet-forming disks influences stellar abundances},
	volume = {676},
	issn = {1432-0746},
	url = {http://dx.doi.org/10.1051/0004-6361/202346604},
	doi = {10.1051/0004-6361/202346604},
	journal = {Astronomy \& Astrophysics},
	author = {H{\"{u}}hn, L.-A. and Bitsch, B.},
	month = aug,
	year = {2023},
	pages = {A87},
}

@article{micolta_vanishing_2025,
	title = {Vanishing {Refractories}: {Tracing} {Dust} {Evolution} in the {BP} {Tau} {Protoplanetary} {Disk}},
	volume = {993},
	issn = {1538-4357},
	url = {http://dx.doi.org/10.3847/1538-4357/ae070c},
	doi = {10.3847/1538-4357/ae070c},
	number = {2},
	journal = {The Astrophysical Journal},
	author = {Micolta, Marbely and Svarc, Justin and Calvet, Nuria and Manzo-Mart{\'{i}}nez, Ezequiel and Thanathibodee, Thanawuth and Magris C., Gladis and Littwiller, Avalon},
	month = nov,
	year = {2025},
	pages = {218},
}

@article{weidenschilling_origin_1997,
	title = {The {Origin} of {Comets} in the {Solar} {Nebula}: {A} {Unified} {Model}},
	volume = {127},
	issn = {0019-1035},
	url = {http://dx.doi.org/10.1006/icar.1997.5712},
	doi = {10.1006/icar.1997.5712},
	number = {2},
	journal = {Icarus},
	author = {Weidenschilling, S.J.},
	month = jun,
	year = {1997},
	pages = {290--306},
}

@article{hartmann_accretion_2016,
	title = {Accretion onto {Pre}-{Main}-{Sequence} {Stars}},
	volume = {54},
	issn = {1545-4282},
	url = {http://dx.doi.org/10.1146/annurev-astro-081915-023347},
	doi = {10.1146/annurev-astro-081915-023347},
	number = {1},
	journal = {Annual Review of Astronomy and Astrophysics},
	author = {Hartmann, Lee and Herczeg, Gregory and Calvet, Nuria},
	month = sep,
	year = {2016},
	pages = {135--180},
}

@article{espaillat_unveiling_2010,
	title = {{UNVEILING} {THE} {STRUCTURE} {OF} {PRE}-{TRANSITIONAL} {DISKS}},
	volume = {717},
	issn = {1538-4357},
	url = {http://dx.doi.org/10.1088/0004-637X/717/1/441},
	doi = {10.1088/0004-637x/717/1/441},
	number = {1},
	journal = {The Astrophysical Journal},
	author = {Espaillat, C. and D'Alessio, P. and Hern{\'{a}}ndez, J. and Nagel, E. and Luhman, K. L. and Watson, D. M. and Calvet, N. and Muzerolle, J. and McClure, M.},
	month = jun,
	year = {2010},
	pages = {441--457},
}

@article{rilinger_determining_2023,
	title = {Determining {Dust} {Properties} in {Protoplanetary} {Disks}: {SED}-derived {Masses} and {Settling} with {ALMA}},
	volume = {944},
	issn = {1538-4357},
	url = {http://dx.doi.org/10.3847/1538-4357/aca905},
	doi = {10.3847/1538-4357/aca905},
	number = {1},
	journal = {The Astrophysical Journal},
	author = {Rilinger, Anneliese M. and Espaillat, Catherine C. and Xin, Zihua and Ribas, {\'{A}}lvaro and Mac{\'{i}}as, Enrique and Luettgen, Sarah},
	month = feb,
	year = {2023},
	pages = {66},
}

@article{mcclure_curved_2013,
	title = {{CURVED} {WALLS}: {GRAIN} {GROWTH}, {SETTLING}, {AND} {COMPOSITION} {PATTERNS} {IN} {T} {TAURI} {DISK} {DUST} {SUBLIMATION} {FRONTS}},
	volume = {775},
	issn = {1538-4357},
	url = {http://dx.doi.org/10.1088/0004-637X/775/2/114},
	doi = {10.1088/0004-637x/775/2/114},
	number = {2},
	journal = {The Astrophysical Journal},
	author = {McClure, M. K. and D'Alessio, P. and Calvet, N. and Espaillat, C. and Hartmann, L. and Sargent, B. and Watson, D. M. and Ingleby, L. and Hern{\'{a}}ndez, J.},
	month = sep,
	year = {2013},
	pages = {114},
}

@article{cardelli_relationship_1989,
	title = {The relationship between infrared, optical, and ultraviolet extinction},
	volume = {345},
	issn = {1538-4357},
	url = {http://dx.doi.org/10.1086/167900},
	doi = {10.1086/167900},
	journal = {The Astrophysical Journal},
	author = {Cardelli, Jason A. and Clayton, Geoffrey C. and Mathis, John S.},
	month = oct,
	year = {1989},
	pages = {245},
}

@article{micolta_ca_2023,
	title = {The {Ca} ii {Lines} as {Tracers} of {Disk} {Structure} in {T} {Tauri} {Stars}: {The} {Chamaeleon} {I} {Region}},
	volume = {953},
	issn = {1538-4357},
	url = {http://dx.doi.org/10.3847/1538-4357/acde52},
	doi = {10.3847/1538-4357/acde52},
	number = {2},
	journal = {The Astrophysical Journal},
	author = {Micolta, Marbely and Calvet, Nuria and Thanathibodee, Thanawuth and Magris C., Gladis and Colmenares, Mar{\'{i}}a Jos{\'{e}} and D{\'{i}}az, Jes{\'{u}}s V. and Alzate-Trujillo, Jairo},
	month = aug,
	year = {2023},
	pages = {177},
}

@article{manoj_spitzer_2011,
	title = {{SPITZER} {INFRARED} {SPECTROGRAPH} {SURVEY} {OF} {YOUNG} {STARS} {IN} {THE} {CHAMAELEON} {I} {STAR}-{FORMING} {REGION}},
	volume = {193},
	issn = {1538-4365},
	url = {http://dx.doi.org/10.1088/0067-0049/193/1/11},
	doi = {10.1088/0067-0049/193/1/11},
	number = {1},
	journal = {The Astrophysical Journal Supplement Series},
	author = {Manoj, P. and Kim, K. H. and Furlan, E. and McClure, M. K. and Luhman, K. L. and Watson, Dan M. and Espaillat, C. and Calvet, N. and Najita, J. R. and D'Alessio, P. and Adame, L. and Sargent, B. A. and Forrest, W. J. and Bohac, C. and Green, J. D. and Arnold, L. A.},
	month = feb,
	year = {2011},
	pages = {11},
}

@article{dalessio_effects_2006,
	title = {Effects of {Dust} {Growth} and {Settling} in {T} {Tauri} {Disks}},
	volume = {638},
	issn = {1538-4357},
	url = {http://dx.doi.org/10.1086/498861},
	doi = {10.1086/498861},
	number = {1},
	journal = {The Astrophysical Journal},
	author = {D'Alessio, Paola and Calvet, Nuria and Hartmann, Lee and Franco-Hernandez, Ramiro and Servin, Hermelinda},
	month = feb,
	year = {2006},
	pages = {314--335},
}

@article{mcclure_observational_2009,
	title = {{OBSERVATIONAL} 5-20 \ensuremath{\mu}m {INTERS}\ensuremath{\mathrm{tel}}{LAR} {EXTINCTION} {CURVES} {TOWARD} {STAR}-{FORMING} {REGIONS} {DERIVED} {FROM} {SPITZER} {IRS} {SPECTRA}},
	volume = {693},
	issn = {1538-4357},
	url = {http://dx.doi.org/10.1088/0004-637X/693/2/L81},
	doi = {10.1088/0004-637x/693/2/l81},
	number = {2},
	journal = {The Astrophysical Journal},
	author = {McClure, M.},
	month = feb,
	year = {2009},
	pages = {L81--L85},
}

@article{pascucci_steeper_2016,
	title = {A {STEEPER} {THAN} {LINEAR} {DISK} {MASS}--{S}\ensuremath{\mathrm{tel}}{LAR} {MASS} {SCALING} {RELATION}},
	volume = {831},
	issn = {1538-4357},
	url = {http://dx.doi.org/10.3847/0004-637X/831/2/125},
	doi = {10.3847/0004-637x/831/2/125},
	number = {2},
	journal = {The Astrophysical Journal},
	author = {Pascucci, I. and Testi, L. and Herczeg, G. J. and Long, F. and Manara, C. F. and Hendler, N. and Mulders, G. D. and Krijt, S. and Ciesla, F. and Henning, Th. and Mohanty, S. and Drabek-Maunder, E. and Apai, D. and Sz{\H{u}}cs, L. and Sacco, G. and Olofsson, J.},
	month = nov,
	year = {2016},
	pages = {125},
}

@article{fedorov_p_vizier_2011,
	title = {{VizieR} {Online} {Data} {Catalog}: {XPM} {Catalog} of positions and proper motions},
	volume = {1319},
	journal = {VizieR Online Data Catalog},
	author = {Fedorov P., Myznikov A., Akhmetov V.},
	year = {2011},
	pages = {I/319--},
}

@ARTICLE{1973A&A....24..337S,
       author = {{Shakura}, N.~I. and {Sunyaev}, R.~A.},
        title = "{Black holes in binary systems. Observational appearance.}",
      journal = {\aap},
         year = 1973,
        month = jan,
       volume = {24},
        pages = {337-355},
       adsurl = {https://ui.adsabs.harvard.edu/abs/1973A&A....24..337S}
}

@misc{dalessio_models_2003,
	howpublished = {arXiv preprint astro-ph/0309590},
	archivePrefix = {arXiv},
	eprint = {astro-ph/0309590},
	title = {Models of {Accretion} {Disks} around {Young} {Stars}},
	url = {http://arxiv.org/pdf/astro-ph/0309590},
	author = {D'Alessio, Paola and Calvet, Nuria and Hartmann, Lee and Muzerolle, James and Sitko, Michael},
	year = {2003},
}

@article{ribas_2017,
    doi = {10.3847/1538-4357/aa8e99},
    url = {http://dx.doi.org/10.3847/1538-4357/aa8e99},
    issn = {1538-4357},
    year = {2017},
    month = {oct},
    pages = {63},
    title = {Far-infrared to Millimeter Data of Protoplanetary Disks: Dust Growth in the
    Taurus, Ophiuchus, and Chamaeleon I Star-forming Regions*},
    author = {Ribas, {\'A}lvaro and Espaillat, Catherine C. and Mac{\'{i}}as, Enrique and Bouy, Herv{\'{e}}
    and Andrews, Sean and Calvet, Nuria and Naylor, David A. and
    Riviere-Marichalar, Pablo and van der Wiel, Matthijs H. D. and Wilner, David},
    number = {1},
    volume = {849},
    journal = {The Astrophysical Journal},
    publisher = {American Astronomical Society}}

@article{luhman_disk_2008,
	title = {The {Disk} {Population} of the {Chamaeleon} {I} {Star}-forming {Region}},
	volume = {675},
	issn = {1538-4357},
	url = {http://dx.doi.org/10.1086/527347},
	doi = {10.1086/527347},
	number = {2},
	journal = {The Astrophysical Journal},
	author = {Luhman, K. L. and Allen, L. E. and Allen, P. R. and Gutermuth, R. A. and Hartmann, L. and Mamajek, E. E. and Megeath, S. T. and Myers, P. C. and Fazio, G. G.},
	month = mar,
	year = {2008},
	pages = {1375--1406},
}

@article{hsieh_campos_2025,
	journal = {\aap},
	volume = {700},
	pages = {A235},
	doi = {10.1051/0004-6361/202555174},
	title = {{CAMPOS} {II}. {The} onset of protostellar disk substructures and planet formation},
	url = {http://arxiv.org/pdf/2504.11577},
	author = {Hsieh, Cheng-Han and Arce, H{\'{e}}ctor G. and Maureira, Mar{\'{i}}a Jos{\'{e}} and Pineda, Jaime E. and Segura-Cox, Dominique and Mardones, Diego and Dunham, Michael M. and Li, Hui and Offner, Stella S. R.},
	year = {2025},
}

@article{gusten_atacama_2006,
	title = {The {Atacama} {Pathfinder} {EXperiment} ({APEX}) -- a new submillimeter facility for southern skies --},
	volume = {454},
	issn = {1432-0746},
	url = {http://dx.doi.org/10.1051/0004-6361:20065420},
	doi = {10.1051/0004-6361:20065420},
	number = {2},
	journal = {Astronomy \& Astrophysics},
	author = {G{\"{u}}sten, R. and Nyman, L. {\r{A}}. and Schilke, P. and Menten, K. and Cesarsky, C. and Booth, R.},
	month = jul,
	year = {2006},
	pages = {L13--L16},
}

@article{siringo_large_2009,
	title = {The {Large} {APEX} {BOlometer} {CAmera} {LABOCA}},
	volume = {497},
	issn = {1432-0746},
	url = {http://dx.doi.org/10.1051/0004-6361/200811454},
	doi = {10.1051/0004-6361/200811454},
	number = {3},
	journal = {Astronomy \& Astrophysics},
	author = {Siringo, G. and Kreysa, E. and Kov{\'{a}}cs, A. and Schuller, F. and Wei{\ss}, A. and Esch, W. and Gem{\"{u}}nd, H.-P. and Jethava, N. and Lundershausen, G. and Colin, A. and G{\"{u}}sten, R. and Menten, K. M. and Beelen, A. and Bertoldi, F. and Beeman, J. W. and Haller, E. E.},
	month = feb,
	year = {2009},
	pages = {945--962},
}

@article{Ochsenbein_2000,
   title={The VizieR database of astronomical catalogues},
   volume={143},
   ISSN={1286-4846},
   url={http://dx.doi.org/10.1051/aas:2000169},
   DOI={10.1051/aas:2000169},
   number={1},
   journal={Astronomy and Astrophysics Supplement Series},
   publisher={EDP Sciences},
   author={Ochsenbein, F. and Bauer, P. and Marcout, J.},
   year={2000},
   month=Apr, pages={23--32} }

@article{belloche_end_2011,
	title = {The end of star formation in {Chamaeleon} {I}?: {A} {LABOCA} census of starless and protostellar cores \ensuremath{\star}\ensuremath{\star}\ensuremath{\star}\ensuremath{\star}\ensuremath{\star}\ensuremath{\star}},
	volume = {527},
	issn = {1432-0746},
	url = {http://dx.doi.org/10.1051/0004-6361/201015733},
	doi = {10.1051/0004-6361/201015733},
	journal = {Astronomy \& Astrophysics},
	author = {Belloche, A. and Schuller, F. and Parise, B. and Andr{\'{e}}, Ph. and Hatchell, J. and J{\o}rgensen, J. K. and Bontemps, S. and Wei{\ss}, A. and Menten, K. M. and Muders, D.},
	month = feb,
	year = {2011},
	pages = {A145},
}

@article{hsieh_alma_2024,
	title = {The {ALMA} {Legacy} {Survey} of {Class} 0/{I} {Disks} in {Corona} australis, {Aquila}, {chaMaeleon}, {oPhiuchus} north, {Ophiuchus}, {Serpens} ({CAMPOS}). {I}. {Evolution} of {Protostellar} {Disk} {Radii}},
	volume = {973},
	url = {https://dx.doi.org/10.3847/1538-4357/ad6152},
	doi = {10.3847/1538-4357/ad6152},
	number = {2},
	journal = {The Astrophysical Journal},
	author = {Hsieh, Cheng-Han and Arce, H{\'{e}}ctor G. and Maureira, Mar{\'{i}}a Jos{\'{e}} and Pineda, Jaime E. and Segura-Cox, Dominique and Mardones, Diego and Dunham, Michael M. and Arun, Aiswarya},
	month = sep,
	year = {2024},
	pages = {138},
}

@article{werner_spitzer_2004,
	title = {The {Spitzer} {Space} {Telescope} {Mission}},
	volume = {154},
	url = {https://dx.doi.org/10.1086/422992},
	doi = {10.1086/422992},
	number = {1},
	journal = {The Astrophysical Journal Supplement Series},
	author = {Werner, M. W. and Roellig, T. L. and Low, F. J. and Rieke, G. H. and Rieke, M. and Hoffmann, W. F. and Young, E. and Houck, J. R. and Brandl, B. and Fazio, G. G. and Hora, J. L. and Gehrz, R. D. and Helou, G. and Soifer, B. T. and Stauffer, J. and Keene, J. and Eisenhardt, P. and Gallagher, D. and Gautier, T. N. and Irace, W. and Lawrence, C. R. and Simmons, L. and Van Cleve, J. E. and Jura, M. and Wright, E. L. and Cruikshank, D. P.},
	month = sep,
	year = {2004},
	pages = {1},
}

@article{houck_infrared_2004,
	title = {The {Infrared} {Spectrograph}* ({IRS}) on the {Spitzer} {Space} {Telescope}},
	volume = {154},
	url = {https://dx.doi.org/10.1086/423134},
	doi = {10.1086/423134},
	number = {1},
	journal = {The Astrophysical Journal Supplement Series},
	author = {Houck, J. R. and Roellig, T. L. and van Cleve, J. and Forrest, W. J. and Herter, T. and Lawrence, C. R. and Matthews, K. and Reitsema, H. J. and Soifer, B. T. and Watson, D. M. and Weedman, D. and Huisjen, M. and Troeltzsch, J. and Barry, D. J. and Bernard-Salas, J. and Blacken, C. E. and Brandl, B. R. and Charmandaris, V. and Devost, D. and Gull, G. E. and Hall, P. and Henderson, C. P. and Higdon, S. J. U. and Pirger, B. E. and Schoenwald, J. and Sloan, G. C. and Uchida, K. I. and Appleton, P. N. and Armus, L. and Burgdorf, M. J. and Fajardo-Acosta, S. B. and Grillmair, C. J. and Ingalls, J. G. and Morris, P. W. and Teplitz, H. I.},
	month = sep,
	year = {2004},
	pages = {18},
}

@article{kobayashi_sublimation_2011,
	title = {Sublimation temperature of circumstellar dust particles and its importance for dust ring formation},
	volume = {63},
	issn = {1880-5981},
	url = {http://dx.doi.org/10.5047/eps.2011.03.012},
	doi = {10.5047/eps.2011.03.012},
	number = {10},
	journal = {Earth, Planets and Space},
	author = {Kobayashi, Hiroshi and Kimura, Hiroshi and Watanabe, Sei-ichiro and Yamamoto, Tetsuo and M{\"{u}}ller, Sebastian},
	month = oct,
	year = {2011},
	pages = {1067--1075},
}

@article{hayashi_gas_2019,
	title = {Gas and {Dust} {Properties} in the {Chamaeleon} {Molecular} {Cloud} {Complex} {Based} on the {Optically} {Thick} {H} i},
	volume = {878},
	url = {https://dx.doi.org/10.3847/1538-4357/ab2051},
	doi = {10.3847/1538-4357/ab2051},
	number = {2},
	journal = {The Astrophysical Journal},
	author = {Hayashi, Katsuhiro and Okamoto, Ryuji and Yamamoto, Hiroaki and Hayakawa, Takahiro and Tachihara, Kengo and Fukui, Yasuo},
	month = jun,
	year = {2019},
	pages = {131},
}

@article{guver_relation_2009,
	title = {The relation between optical extinction and hydrogen column density in the {Galaxy}},
	volume = {400},
	issn = {0035-8711},
	url = {https://doi.org/10.1111/j.1365-2966.2009.15598.x},
	doi = {10.1111/j.1365-2966.2009.15598.x},
	number = {4},
	journal = {Monthly Notices of the Royal Astronomical Society},
	author = {G{\"{u}}ver, Tolga and {\"{O}}zel, Feryal},
	month = dec,
	year = {2009},
	pages = {2050--2053},
}

@article{lodders_solar_2003,
	title = {Solar {System} {Abundances} and {Condensation} {Temperatures} of the {Elements}},
	volume = {591},
	issn = {1538-4357},
	url = {http://dx.doi.org/10.1086/375492},
	doi = {10.1086/375492},
	number = {2},
	journal = {The Astrophysical Journal},
	author = {Lodders, Katharina},
	month = jul,
	year = {2003},
	pages = {1220--1247},
}

@article{flaherty_infrared_2007,
	title = {Infrared {Extinction} toward {Nearby} {Star}-forming {Regions}},
	volume = {663},
	issn = {1538-4357},
	url = {http://dx.doi.org/10.1086/518411},
	doi = {10.1086/518411},
	number = {2},
	journal = {The Astrophysical Journal},
	author = {Flaherty, K. M. and Pipher, J. L. and Megeath, S. T. and Winston, E. M. and Gutermuth, R. A. and Muzerolle, J. and Allen, L. E. and Fazio, G. G.},
	month = jul,
	year = {2007},
	pages = {1069--1082},
}

@article{calvet_structure_1998,
	title = {The {Structure} and {Emission} of the {Accretion} {Shock} in {T} {Tauri} {Stars}},
	volume = {509},
	issn = {1538-4357},
	url = {http://dx.doi.org/10.1086/306527},
	doi = {10.1086/306527},
	number = {2},
	journal = {The Astrophysical Journal},
	author = {Calvet, Nuria and Gullbring, Erik},
	month = dec,
	year = {1998},
	pages = {802--818},
}

@article{chapman_mid-infrared_2008,
	title = {{THE} {MID}-{INFRARED} {EXTINCTION} {LAW} {IN} {THE} {OPHIUCHUS}, {PERSEUS}, {AND} {SERPENS} {MOLECULAR} {CLOUDS}},
	volume = {690},
	issn = {1538-4357},
	url = {http://dx.doi.org/10.1088/0004-637X/690/1/496},
	doi = {10.1088/0004-637x/690/1/496},
	number = {1},
	journal = {The Astrophysical Journal},
	author = {Chapman, Nicholas L. and Mundy, Lee G. and Lai, Shih-Ping and Evans, Neal J.},
	month = dec,
	year = {2008},
	pages = {496--511},
}

@article{mauco_herschel_2018,
	title = {Herschel {PACS} {Observations} of 4--10 {Myr} {Old} {Classical} {T} {Tauri} {Stars} in {Orion} {OB1}},
	volume = {859},
	issn = {1538-4357},
	url = {http://dx.doi.org/10.3847/1538-4357/aabf40},
	doi = {10.3847/1538-4357/aabf40},
	number = {1},
	journal = {The Astrophysical Journal},
	author = {Mauc{\'{o}}, Karina and Brice{\~{n}}o, C{\'{e}}sar and Calvet, Nuria and Hern{\'{a}}ndez, Jes{\'{u}}s and Ballesteros-Paredes, Javier and Gonz{\'{a}}lez, Omaira and C. Espaillat, Catherine and Li, Dan and Telesco, Charles M. and Downes, Juan Jos{\'{e}} and Mac{\'{i}}as, Enrique and Qi, Chunhua and Michel, Ra{\'{u}}l and D'Alessio, Paola and Ali, Babar},
	month = may,
	year = {2018},
	pages = {1},
}

@article{luhman_stellar_2007,
	title = {The {Stellar} {Population} of the {Chamaeleon} {I} {Star}-forming {Region}},
	volume = {173},
	issn = {1538-4365},
	url = {http://dx.doi.org/10.1086/520114},
	doi = {10.1086/520114},
	number = {1},
	journal = {The Astrophysical Journal Supplement Series},
	author = {Luhman, K. L.},
	month = nov,
	year = {2007},
	pages = {104--136},
}

@article{brogan_2014_2015,
	title = {{THE} 2014 {ALMA} {LONG} {BASELINE} {CAMPAIGN}: {FIRST} {RESULTS} {FROM} {HIGH} {ANGULAR} {RESOLUTION} {OBSERVATIONS} {TOWARD} {THE} {HL} {TAU} {REGION}},
	volume = {808},
	issn = {2041-8213},
	url = {http://dx.doi.org/10.1088/2041-8205/808/1/L3},
	doi = {10.1088/2041-8205/808/1/l3},
	number = {1},
	journal = {The Astrophysical Journal},
	author = {Brogan, C. L. and P{\'{e}}rez, L. M. and Hunter, T. R. and Dent, W. R. F. and Hales, A. S. and Hills, R. E. and Corder, S. and Fomalont, E. B. and Vlahakis, C. and Asaki, Y. and Barkats, D. and Hirota, A. and Hodge, J. A. and Impellizzeri, C. M. V. and Kneissl, R. and Liuzzo, E. and Lucas, R. and Marcelino, N. and Matsushita, S. and Nakanishi, K. and Phillips, N. and Richards, A. M. S. and Toledo, I. and Aladro, R. and Broguiere, D. and Cortes, J. R. and Cortes, P. C. and Espada, D. and Galarza, F. and Appadoo, D. Garcia- and Ramirez, L. Guzman- and Humphreys, E. M. and Jung, T. and Kameno, S. and Laing, R. A. and Leon, S. and Marconi, G. and Mignano, A. and Nikolic, B. and Nyman, L.-A. and Radiszcz, M. and Remijan, A. and Rod{\'{o}}n, J. A. and Sawada, T. and Takahashi, S. and Tilanus, R. P. J. and Vilaro, B. Vila and Watson, L. C. and Wiklind, T. and Akiyama, E. and Chapillon, E. and Monsalvo, I. de Gregorio- and Francesco, J. Di and Gueth, F. and Kawamura, A. and Lee, C.-F. and Luong, Q. Nguyen and Mangum, J. and Pietu, V. and Sanhueza, P. and Saigo, K. and Takakuwa, S. and Ubach, C. and Kempen, T. van and Wootten, A. and Carrizo, A. Castro- and Francke, H. and Gallardo, J. and Garcia, J. and Gonzalez, S. and Hill, T. and Kaminski, T. and Kurono, Y. and Liu, H.-Y. and Lopez, C. and Morales, F. and Plarre, K. and Schieven, G. and Testi, L. and Videla, L. and Villard, E. and Andreani, P. and Hibbard, J. E. and Tatematsu, K.},
	month = jul,
	year = {2015},
	pages = {L3},
}

@article{kelson_optimal_2003,
	title = {Optimal {Techniques} in {Two}-dimensional {Spectroscopy}: {Background} {Subtraction} for the 21st {Century}},
	volume = {115},
	issn = {1538-3873},
	url = {http://dx.doi.org/10.1086/375502},
	doi = {10.1086/375502},
	number = {808},
	journal = {Publications of the Astronomical Society of the Pacific},
	author = {Kelson, Daniel D.},
	month = jun,
	year = {2003},
	pages = {688--699},
}

@ARTICLE{Frasca2006,
       author = {{Frasca}, A. and {Guillout}, P. and {Marilli}, E. and {Freire Ferrero}, R. and {Biazzo}, K. and {Klutsch}, A.},
        title = "{Newly discovered active binaries in the RasTyc sample of stellar X-ray sources. I. Orbital and physical parameters of six new binaries}",
      journal = {\aap},
         year = 2006,
        month = jul,
       volume = {454},
       number = {1},
        pages = {301-309},
          doi = {10.1051/0004-6361:20054573},
       adsurl = {https://ui.adsabs.harvard.edu/abs/2006A&A...454..301F}
}

@ARTICLE{Randich2022,
       author = {{Randich}, S. and {Gilmore}, G. and {Magrini}, L. and {Sacco}, G.~G. and {Jackson}, R.~J. and {Jeffries}, R.~D. and {Worley}, C.~C. and {Hourihane}, A. and {Gonneau}, A. and {Viscasillas Vazquez}, C. and {Franciosini}, E. and {Lewis}, J.~R. and {Alfaro}, E.~J. and {Allende Prieto}, C. and {Bensby}, T. and {Blomme}, R. and {Bragaglia}, A. and {Flaccomio}, E. and {Fran{\c{c}}ois}, P. and {Irwin}, M.~J. and {Koposov}, S.~E. and {Korn}, A.~J. and {Lanzafame}, A.~C. and {Pancino}, E. and {Recio-Blanco}, A. and {Smiljanic}, R. and {Van Eck}, S. and {Zwitter}, T. and {Asplund}, M. and {Bonifacio}, P. and {Feltzing}, S. and {Binney}, J. and {Drew}, J. and {Ferguson}, A.~M.~N. and {Micela}, G. and {Negueruela}, I. and {Prusti}, T. and {Rix}, H. -W. and {Vallenari}, A. and {Bayo}, A. and {Bergemann}, M. and {Biazzo}, K. and {Carraro}, G. and {Casey}, A.~R. and {Damiani}, F. and {Frasca}, A. and {Heiter}, U. and {Hill}, V. and {Jofr{\'e}}, P. and {de Laverny}, P. and {Lind}, K. and {Marconi}, G. and {Martayan}, C. and {Masseron}, T. and {Monaco}, L. and {Morbidelli}, L. and {Prisinzano}, L. and {Sbordone}, L. and {Sousa}, S.~G. and {Zaggia}, S. and {Adibekyan}, V. and {Bonito}, R. and {Caffau}, E. and {Daflon}, S. and {Feuillet}, D.~K. and {Gebran}, M. and {Gonzalez Hernandez}, J.~I. and {Guiglion}, G. and {Herrero}, A. and {Lobel}, A. and {Maiz Apellaniz}, J. and {Merle}, T. and {Mikolaitis}, {\v{S}}. and {Montes}, D. and {Morel}, T. and {Soubiran}, C. and {Spina}, L. and {Tabernero}, H.~M. and {Tautvai{\v{s}}iene}, G. and {Traven}, G. and {Valentini}, M. and {Van der Swaelmen}, M. and {Villanova}, S. and {Wright}, N.~J. and {Abbas}, U. and {Aguirre B{\o}rsen-Koch}, V. and {Alves}, J. and {Balaguer-Nunez}, L. and {Barklem}, P.~S. and {Barrado}, D. and {Berlanas}, S.~R. and {Binks}, A.~S. and {Bressan}, A. and {Capuzzo-Dolcetta}, R. and {Casagrande}, L. and {Casamiquela}, L. and {Collins}, R.~S. and {D'Orazi}, V. and {Dantas}, M.~L.~L. and {Debattista}, V.~P. and {Delgado-Mena}, E. and {Di Marcantonio}, P. and {Drazdauskas}, A. and {Evans}, N.~W. and {Famaey}, B. and {Franchini}, M. and {Fr{\'e}mat}, Y. and {Friel}, E.~D. and {Fu}, X. and {Geisler}, D. and {Gerhard}, O. and {Gonzalez Solares}, E.~A. and {Grebel}, E.~K. and {Gutierrez Albarran}, M.~L. and {Hatzidimitriou}, D. and {Held}, E.~V. and {Jim{\'e}nez-Esteban}, F. and {J{\"o}nsson}, H. and {Jordi}, C. and {Khachaturyants}, T. and {Kordopatis}, G. and {Kos}, J. and {Lagarde}, N. and {Mahy}, L. and {Mapelli}, M. and {Marfil}, E. and {Martell}, S.~L. and {Messina}, S. and {Miglio}, A. and {Minchev}, I. and {Moitinho}, A. and {Montalban}, J. and {Monteiro}, M.~J.~P.~F.~G. and {Morossi}, C. and {Mowlavi}, N. and {Mucciarelli}, A. and {Murphy}, D.~N.~A. and {Nardetto}, N. and {Ortolani}, S. and {Paletou}, F. and {Palou{\v{s}}}, J. and {Paunzen}, E. and {Pickering}, J.~C. and {Quirrenbach}, A. and {Re Fiorentin}, P. and {Read}, J.~I. and {Romano}, D. and {Ryde}, N. and {Sanna}, N. and {Santos}, W. and {Seabroke}, G.~M. and {Spagna}, A. and {Steinmetz}, M. and {Stonkut{\'e}}, E. and {Sutorius}, E. and {Th{\'e}venin}, F. and {Tosi}, M. and {Tsantaki}, M. and {Vink}, J.~S. and {Wright}, N. and {Wyse}, R.~F.~G. and {Zoccali}, M. and {Zorec}, J. and {Zucker}, D.~B. and {Walton}, N.~A.},
        title = "{The Gaia-ESO Public Spectroscopic Survey: Implementation, data products, open cluster survey, science, and legacy}",
      journal = {\aap},
         year = 2022,
        month = oct,
       volume = {666},
          eid = {A121},
        pages = {A121},
          doi = {10.1051/0004-6361/202243141},
archivePrefix = {arXiv},
       eprint = {2206.02901},
 primaryClass = {astro-ph.GA},
       adsurl = {https://ui.adsabs.harvard.edu/abs/2022A&A...666A.121R}
}

@ARTICLE{Moultaka2004,
       author = {{Moultaka}, J. and {Ilovaisky}, S.~A. and {Prugniel}, P. and {Soubiran}, C.},
        title = "{The ELODIE Archive}",
      journal = {\pasp},
         year = 2004,
        month = jul,
       volume = {116},
       number = {821},
        pages = {693-698},
          doi = {10.1086/422177},
archivePrefix = {arXiv},
       eprint = {astro-ph/0405025},
 primaryClass = {astro-ph},
       adsurl = {https://ui.adsabs.harvard.edu/abs/2004PASP..116..693M}
}

@ARTICLE{Frasca2015,
       author = {{Frasca}, A. and {Biazzo}, K. and {Lanzafame}, A.~C. and {Alcal{\'a}}, J.~M. and {Brugaletta}, E. and {Klutsch}, A. and {Stelzer}, B. and {Sacco}, G.~G. and {Spina}, L. and {Jeffries}, R.~D. and {Montes}, D. and {Alfaro}, E.~J. and {Barentsen}, G. and {Bonito}, R. and {Gameiro}, J.~F. and {L{\'o}pez-Santiago}, J. and {Pace}, G. and {Pasquini}, L. and {Prisinzano}, L. and {Sousa}, S.~G. and {Gilmore}, G. and {Randich}, S. and {Micela}, G. and {Bragaglia}, A. and {Flaccomio}, E. and {Bayo}, A. and {Costado}, M.~T. and {Franciosini}, E. and {Hill}, V. and {Hourihane}, A. and {Jofr{\'e}}, P. and {Lardo}, C. and {Maiorca}, E. and {Masseron}, T. and {Morbidelli}, L. and {Worley}, C.~C.},
        title = "{The Gaia-ESO Survey: Chromospheric emission, accretion properties, and rotation in {\ensuremath{\gamma}} Velorum and Chamaeleon I{\ensuremath{\star}}{\ensuremath{\star}}{\ensuremath{\star}}}",
      journal = {\aap},
         year = 2015,
        month = mar,
       volume = {575},
          eid = {A4},
        pages = {A4},
          doi = {10.1051/0004-6361/201424409},
archivePrefix = {arXiv},
       eprint = {1412.4153},
 primaryClass = {astro-ph.SR},
       adsurl = {https://ui.adsabs.harvard.edu/abs/2015A&A...575A...4F}
}
\bibliographystyle{aasjournalv7}

\end{document}